\documentclass[JHEP,11pt]{article}
\usepackage{verbatim}

\usepackage{dsfont}
\usepackage{sidecap}
\sidecaptionvpos{figure}{t}
\usepackage{subfigure}
\usepackage[utf8]{inputenc}
\usepackage[english]{babel}
\usepackage{empheq}
\usepackage{amssymb,amsmath}
\usepackage{graphicx, xcolor, varwidth}
\usepackage{setspace}
\usepackage[permil]{overpic}
\usepackage{cite}
\usepackage{graphicx}% see geometry.pdf on how to lay out the page. There's lots.
\usepackage{color}
\usepackage{braket}
\usepackage{simplewick}
\usepackage{jheppub}
\usepackage[T1]{fontenc}

\usepackage[bottom]{footmisc}

\usepackage{simpler-wick}

\usepackage{feynmp}
\colorlet{darkblue}{blue!70!black}
\colorlet{darkgreen}{green!70!black}

\numberwithin{equation}{section}

\newcommand{\be}{\begin{equation}}
\newcommand{\ee}{\end{equation}}
\newcommand{\bea}{\begin{eqnarray}}
\newcommand{\eea}{\end{eqnarray}}
\newcommand{\bear}{\begin{eqnarray}}
\newcommand{\eear}{\end{eqnarray}}  
\newcommand{\beas}{\begin{eqnarray*}}
	
	\newcommand{\eeas}{\end{eqnarray*}}
\newcommand{\ba}{\begin{array}}
	\newcommand{\ea}{\end{array}}

\def\ba#1\ea{\begin{align}#1\end{align}}

\newcommand{\gone}{\gamma^{(1)}}
\newcommand{\cone}{c^{(1)}}
\newcommand{\ctwo}{c^{(2)}}

\newcommand{\tr}{\operatorname{Tr}}
\newcommand{\pd}[2][1]{\ifnum#1=1 \frac{\partial}{\partial {#2}} \else
	\frac{\partial^#1}{\partial {#2}^{#1}}\fi}
\newcommand{\dpd}[2][1]{\ifnum#1=1 \dfrac{\partial}{\partial {#2}} \else
	\frac{\partial^#1}{\partial {#2}^{#1}}\fi}
\newcommand{\td}[2][1]{\ifnum#1=1 \frac{d}{d{#2}} \else
	\frac{d^#1}{d{#2}^{#1}}\fi}

\newcommand{\col}{\mathop{:}}
\newcommand{\nbox}{{\,\lower0.9pt\vbox{\hrule \hbox{\vrule height 0.2 cm \hskip 0.19 cm \vrule height 0.2 cm}\hrule}\,}}

\newcommand{\bz}{\bar{z}}

\newcommand{\bw}{\bar{w}}

\renewcommand{\Im}{\mbox{Im\ }}



		\title{	{ Entropy Variations and Light Ray Operators from Replica Defects }}

			\author[a]{Srivatsan Balakrishnan} 
			\author[b]{Venkatesa Chandrasekaran} 
			\author[a]{Thomas Faulkner}
			\author[b]{Adam Levine}
			\author[b]{Arvin Shahbazi-Moghaddam}

		\affiliation[a]{Department of Physics, University of Illinois, 1110 W. Green St., Urbana IL 61801-3080, U.S.A}
		\affiliation[b]{Center for Theoretical Physics and Department of Physics, University of California, Berkeley, CA 94720}
		
		\emailAdd{sblkrsh2@illinois.edu}
		\emailAdd{ven\_chandrasekaran@berkeley.edu}
		\emailAdd{tomf@illinois.edu, arlevine@berkeley.edu}
		\emailAdd{arlevine@berkeley.edu}
		\emailAdd{arvinshm@berkeley.edu}
			
		\abstract{
			We study the defect operator product expansion (OPE) of displacement operators in free and interacting conformal field theories using replica methods. We show that as $n$ approaches $1$ a contact term can emerge when the OPE  contains defect operators of twist $d-2$. For interacting theories and general states we give evidence  that the only possibility is from the defect operator that becomes the stress tensor in the $n\to 1$ limit. 
			This implies that the quantum null energy condition (QNEC) is always saturated for CFTs with a twist gap. As a check, we show independently that in a large class of near vacuum states, the second variation of the entanglement entropy is given by a simple correlation function of averaged null energy operators as studied by Hofman and Maldacena. 
			This suggests that sub-leading terms in the the defect OPE are controlled by a defect version of the spin-3 non-local light ray operator and we speculate about the possible origin of such a defect operator. For free theories this contribution condenses to a contact term that leads to violations of QNEC saturation.
	}

\addtocounter{page}{1}

\begin{document}
	\maketitle 
\section{Introduction}
Despite much progress in understanding entanglement entropy using bulk geometric methods in holographic field theories \cite{Ryu:2006bv,Ryu:2006ef,Hubeny:2007xt}, significantly less progress has been made on the more difficult problem of computing entanglement entropy directly in field theory. Part of what makes entanglement entropy such a difficult object to study in field theory is its inherently non-local and state-dependent nature.

One way to access the structure of entanglement in field theories is to study its dependence on the shape of the entangling surface. Such considerations have led to important results on the nature of entanglement in quantum field theories \cite{Faulkner:2016mzt, Bousso:2015wca,Koeller:2015qmn, Akers:2017aa, Bousso:2016aa, Balakrishnan:2017aa, Allais:2014ata, Faulkner:2016aa, Mezei:2015aa}. To study the shape dependence of entanglement entropy for QFTs in $d>2$ dimensions,
consider a Cauchy slice $\Sigma$ containing a subregion $\mathcal{R}$ with entangling surface $\partial \mathcal{R}$ in a general conformal field theory. By unitary equivalence of Cauchy slices which intersect the same surface $\partial \mathcal{R}$, the entanglement entropy for some fixed global state can be viewed as a functional of the entangling surface embedding coordinates $X^{\mu}(y^i)$ where the $y^i$ with $i = 1,...,d-2$ are internal coordinates on $\partial \mathcal{R}$. We write:

\begin{align}
S_{\mathcal{R}} = S[X(y)].
\end{align}
The shape dependence of the entanglement entropy can then be accessed by taking functional derivatives. In particular, we can expand the entanglement entropy about some background entangling surface $X(y) = X_0(y) + \delta X(y)$ as 
\begin{align}
S[X] &= S[X_0] + \int d^{d-2}y \left. \frac{\delta S_{\mathcal{R}}}{\delta X^{\mu}(y)} \right \vert_{X_0} \delta X^{\mu}(y) 
\nonumber \\ &+ \int d^{d-2} y d^{d-2} y' \left. \frac{\delta^2 S_{\mathcal{R}}}{\delta X^{\mu}(y) \delta X^{\nu}(y')} \right \vert_{X_0} \delta X^{\mu}(y) \delta X^{\nu}(y') + ...\ .
\end{align}

\begin{figure}[]
\center \includegraphics[width=.75\textwidth]{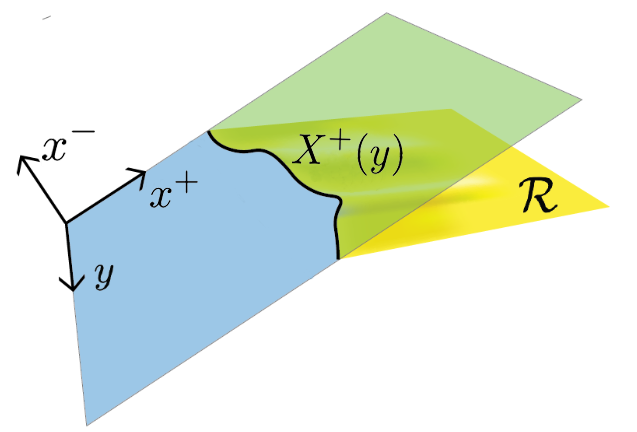}
\caption{We consider the entanglement entropy associated to a spatial subregion $\mathcal{R}$. The entangling surface lies along $x^- = 0$ and $x^+ = X^+(y)$. In this work, we study the dependence of the entanglement entropy on the profile $X^+(y)$.}
\label{fig1}
\end{figure}

This second variation has received a lot of attention in part because it is an essential ingredient in defining the \emph{quantum null energy condition} (QNEC) \cite{Bousso:2015mna,Bousso:2016aa}. The QNEC bounds the null-null component of the stress tensor at a point by a specific contribution from the second shape variation of the entanglement entropy. More specifically, this second variation can be naturally split into two pieces - the \emph{diagonal} term which is proportional to a delta function in the internal coordinates $y^i$ and the \emph{off-diagonal} terms\footnote{Note that the entanglement entropy, being UV divergent, will typically have divergent contributions that are local to the entangling surface. These will show up as a limited set of diagonal/contact terms in \eqref{eqn:secondvar}. For deformations about a sufficiently flat entangling surface these terms do not contribute to
the contact term that is the subject of the QNEC. The divergent terms will not be the subject of investigation here.}
\begin{align}\label{eqn:secondvar}
\frac{\delta^2 S_\mathcal{R}}{\delta X^{+}(y) \delta X^{+}(y')} = S^{''}(y) \delta^{(d-2)}(y-y') + (\text{off-diagonal}). 
\end{align}
where $(X^+, X^-)$ are the null directions orthogonal to the defect. The QNEC states that the null energy flowing past a point must be lower bounded by the diagonal second variation
\begin{align}\label{eqn:QNEC}
\braket{T_{++}(y)} \geq \frac{\hbar}{2\pi} S^{''}(y) ,
\end{align} 
where we are taking $\mathcal{R}$ to be a Rindler wedge. This inequality was first proposed as the $G_N \to 0$ limit of the quantum focussing conjecture \cite{Bousso:2015mna}, and was first proven in free and super-renormalizable field theories in \cite{Bousso:2015wca}. The proof for general QFTs with an interacting UV fixed point was given in \cite{Balakrishnan:2017aa}. More recently, yet another proof was given using techniques from algebraic quantum field theory \cite{Ceyhan:2018zfg}. 

The method of proof in the free case involved explicitly computing $S_{++}^{''}$ where it was found that
\begin{align}
S^{''} = \frac{2\pi}{\hbar} \braket{T_{++}} - Q
\end{align} where for general states $Q \geq 0$. In contrast, the proof in general QFTs relied on relating the inequality \eqref{eqn:QNEC} to the causality of a certain correlation function involving modular flow. 
 This left open the question of whether $S^{''}$ could be explicility computed in more general field theories. 
 
In \cite{Leichenauer:2018obf} the diagonal term $S^{''}$ was  computed in large $N$ QFTs in states with a geometric dual. Remarkably, the result was 
\begin{align}\label{eqn:qnecsaturation}
S^{''}(y) = 2\pi \braket{T_{++}(y)}  
\end{align}
where we have now set $\hbar = 1$. In other words, $Q=0$ for such theories.  In that work, it was argued that neither finite coupling nor finite $N$ corrections should affect this formula. This led the authors of \cite{Leichenauer:2018obf} to conjecture \eqref{eqn:qnecsaturation} for all interacting CFTs. The main goal of this paper is to provide evidence for \eqref{eqn:qnecsaturation} in general CFTs with a twist gap.

The method of argument will follow from the replica trick for computing entanglement entropy. The replica trick uses the formula
\begin{align}
S[\mathcal{R}] = \lim_{n\to 1} (1-n\partial_n)\log \tr[\rho_{\mathcal{R}}^n]
\end{align} 
to relate the entanglement entropy to the partition function of the CFT on a replicated manifold \cite{Holzhey:1994we,Calabrese:2004eu} (see also \cite{Lashkari:2015dia, Sarosi:2016oks, Ruggiero:2016khg,Lewkowycz:2013nqa})
\begin{align}
\tr[\rho^n_{\mathcal{R}}] = Z_n/(Z_1)^n.
\end{align}
At integer $n$, $Z_n$ can be computed via a path integral on a branched manifold with $n$-sheets. Alternatively, one can compute this as a path integral on an unbranched manifold but in the presence of a twist defect operator $\Sigma_n$ of co-dimension 2 that lives at the entangling surface \cite{Bianchi_2016}. Doing so allows us to employ techniques from defect CFTs. See \cite{Billo:2016aa,Gliozzi:2015qsa,Gaiotto:2013nva, Billo:2013jda} for a general introduction to these tools. 

In particular, shape deformations of the defect are controlled by a defect operator, namely the displacement operator, with components $\hat{D}_{+}, \hat{D}_{-}$. This operator is universal to defect CFTs. Its importance in entanglement entropy computations was elucidated in \cite{Billo:2016aa,Balakrishnan:2017aa, Bianchi_2016}. Consequently, the second variation of the entanglement entropy is related to the two-point function of displacement operators 
\begin{align}
\frac{\delta^2 S}{\delta X^{+}(y) \delta X^{+}(y')} = \lim_{n\to 1} \frac{-2\pi}{n-1} \braket{\Sigma_n^{\psi} \hat{D}_{+}(y) \hat{D}_{+}(y')},
\end{align}
where the notation $\Sigma_n^{\psi}$ will be explained in the next section.

Since we are interested in the delta function contribution to this second variation, we can take the limit where the two displacement operators approach each other, $y \to y'$. This suggests that we should study the OPE of two displacement operators and look for terms which produce a delta function, at least as $n \to 1$.

 It might seem strange to look for a delta function in an OPE since the latter, without further input, results in an expansion in powers of $|y-y'|$. We will find a delta function can emerge from a delicate interplay between the OPE and the replica limit $n \rightarrow 1$.

An obvious check of our understanding of \eqref{eqn:qnecsaturation} is to explain how this formula can be true for interacting theories while there exist states for which $Q>0$ in free theories. This is a particularly pertinent concern in, for example, $\mathcal{N}=4$ super-Yang Mills where one can tune the coupling to zero while remaining at a CFT fixed point. We will find that in the free limit certain terms in the off-diagonal contributions of \eqref{eqn:secondvar} become more singular and ``condense" into a delta function in the zero coupling limit. In a weakly interacting theory it becomes a question of resolution as to whether one considers $Q$ to be zero or not. 

In fact this phenomenon is not unprecedented. The authors of \cite{Hofman_2008} studied energy correlation functions in a so called conformal collider setup.  The statistical properties of the angular distribution of energy in excited states collected at long distances is very different for free and interacting CFTs. We conjecture that these situations are controlled by the same physics.
Explicitly, in certain special ``near vacuum'' states, there is a contribution to the second variation of entanglement that can be written in terms of these energy correlation functions.

Schematically, we will find
\begin{align}\label{eqn:E+E+}
 \frac{\delta^2 S}{\delta X^{+}(y) \delta X^{+}(y')}  - \frac{2\pi}{\hbar} \braket {T_{++}} \delta^{(d-2)} (y-y')  \sim \int ds e^s \braket{\mathcal{O} \hat{\mathcal{E}}_+(y) \hat{\mathcal{E}}_+(y')e^{iKs} \mathcal{O}}
\end{align}
where 
\begin{align}
\hat{\mathcal{E}}_+(y) = \int_{-\infty}^{\infty} d\lambda \braket{T_{++}(x^+ = \lambda,x^-=0,y)}
\end{align}
is the averaged null energy operator discussed in \cite{Hofman_2008} and the $\mathcal{O}$'s should be thought of as state-creation operators. The operator $K$ is the boost generator about the undeformed entangling surface. 

The singularities in $|y-y'|$ of the correlator in \eqref{eqn:E+E+} are then understood by taking the OPE of two averaged null energy operators. This OPE was first discussed in \cite{Hofman_2008} where a new non-local ``light ray" operator of spin 3 was found to control the small $y-y'$ limit.

In the free limit, we will show that this non-local operator has the correct scaling dimension to give rise to a new delta function term in \eqref{eqn:E+E+}. In the interacting case this operator picks up an anomalous dimension and thus lifts the delta function.

In other words, the presence of an extra delta function in the second variation of the entanglement entropy in free theories can be viewed as a manifestation of the singular behavior of the conformal collider energy correlation functions in free theories. This is just another manifestation of the important relationship between entanglement and energy density in QFT.

The presence of this spin-3 light ray operator in the shape variation of entanglement
in specific states however points to an issue with our defect OPE argument. In particular one can show that this contribution cannot come directly from one of the local defect operators that we enumerated in order to argue for saturation.  Thus one might worry that there are other additional non-trivial contributions to the OPE that we miss by simply analyzing this local defect spectrum. The main issue seems to be that the $n \rightarrow 1$ limit does not commute with the OPE limit. Thus in order to take the limit in the proper order we should first re-sum a subset of the defect operators in the OPE before taking the limit $n \rightarrow 1$. 
For specific states we can effectively achieve this resummation (by giving a general expression valid for finite $|y-y'|$) however for general states we have not managed to do this. Thus, we are not sure how this spin-3 light ray operator will show up for more general states beyond those covered by \eqref{eqn:E+E+}. Nevertheless we will refer to these non-standard contributions as arising from ``nonlocal defect operators.''

The basic reason it is hard to make a general statement is that entanglement can be thought of as a state dependent observable. This state dependence shows up in the replica trick as a non-trivial $n$ dependence in the limit $n \rightarrow 1$ so the order of limits issue discussed above is linked to this state dependence. 
We are thus left to compute the OPE of two displacement operators for some specific states and configurations. This allows us to check the power laws that appear in the $|y_1-y_2|$ expansion for possible saturation violations. 
Given this we present two main pieces of evidence that the nonlocal defect operators do not lead to violations of QNEC saturation. The first is the aforementioned near vacuum state calculation. The second is a new calculation of the fourth shape variation of \emph{vacuum} entanglement entropy which is also sensitive to the displacement operator defect OPE. In both cases we find that the only new operator that shows up is the spin-3 light ray operator. 

\noindent The outline of the paper is as follows.
\begin{itemize}
\item In Section \ref{sec:background}, we begin by reviewing the basics of the replica trick and the relevant ideas from defect conformal field theory. We review the spectrum of local operators that are induced on the defect, including the infinite family of so-called higher spin displacement operators. We show that, in an interacting theory, these higher spin operators by themselves cannot contribute to the diagonal QNEC. We also present a present a certain conjecture about the nonlocal defect operators.

\item In Section \ref{sec:saturation}, we discuss how a delta function appears in the OPE of two displacement operators. We focus on a specific defect operator that limits to $T_{++}$ as $n \rightarrow 1$.
For this defect operator we derive a prediction for the ratio of the $D_+ D_+$ OPE coefficient and its anomalous defect dimension. In Section \ref{sec:Tpp}, we check this prediction by making use of a modified Ward identity for the defect theory. In Appendix~\ref{cn}-\ref{gn} we also explicitly compute the anomalous dimension and the OPE coefficient to confirm this prediction.

\item 
In Section~\ref{sec:proof}, we take up the concern that there could be other operators which lead to delta functions even for interacting CFTs. To do this, we compute the defect four point function $\mathcal{F}_n:=\braket{\Sigma_n^0 \hat{D}_+(y_1) \hat{D}_+(y_2) \hat{D}_-(y_3) \hat{D}_-(y_4)}$ in the limit $n \rightarrow 1$. From this we can read off the spectrum by analyzing the powers of $|y_1- y_2|$ that appear as $y_1 \rightarrow y_2$. We will find that these powers arise from the light-ray OPE of two averaged null energy operators. 

\item Finally, in Section \ref{sec:nearvacuum}, we check our results by explicitly computing the entanglement entropy second variation in near-vacuum states. By using null quantiation for free theories, we show that our results agree with that of \cite{Bousso:2016aa}. 
\item In Section \ref{sec:Discussion}, we end with a discussion of our results. 
\end{itemize}

\section{Replica Trick and the Displacement Operator}\label{sec:background}

In this section, we will review the replica trick and discuss the connection between entanglement entropy and defect operators. This naturally leads to the displacement operator, which will be the key tool for studying \eqref{eqn:qnecsaturation}. 

As outlined in the introduction, the replica trick instructs us to compute the partition function $Z_n/(Z_1)^n = \tr[\rho_{\mathcal{R}}^n]$, which can be understood as a path integral on a branched manifold $\mathcal{M}_n(\mathcal{R})$, where taking the product of density matrices acts to glue each consecutive sheet together. Using the state operator correspondence, a general state can be represented by the insertion of of a scalar operator in the Euclidean section, so that 
\begin{align}
Z_n = \braket{\psi^{\dagger \otimes n} \psi^{\otimes n}}_{\mathcal{M}_n(\mathcal{R})}
\end{align}
where each $\psi$ is inserted on cyclicly consecutive sheets. Alternatively, we can view this $2n$-point correlation function as being computed not on an $n$-sheeted manifold but on a manifold with trivial topology in the presence of a codimension 2 twist defect operator
\begin{align}
Z_n =  \braket{\Sigma_n^0\psi^{\dagger \otimes n} \psi^{\otimes n}}_{\text{CFT}^{\otimes n}/\mathbb{Z}_n} \equiv \braket{\Sigma_n^{\psi}}
\end{align}
where we have used a compact notation for the twist operator that includes the state operator insertions: $\Sigma_n^{\psi} \equiv\Sigma_n^0\psi^{\dagger \otimes n} \psi^{\otimes n}$. It is convenient (and possible) to orbifold the $\text{CFT}^{\otimes n}$ which projects onto states in the singlet of $\mathbb{Z}_n$. This allows us to work with a CFT that for example has only one conserved stress tensor. 

We take the defect $\Sigma_n^0$ to be associated to a flat cut of a null plane in Minkowski space. We take the metric to be 
\begin{align}
ds^2 = dz d\bz + d\vec{y}^2
\end{align}
where $z$ and $\bz$ are complexified lightcone coordinates. That is, on the Lorentzian section we have $z = -x^-= x+i\tau$ and $\bz = x^+ = x-i\tau$. Thus, we take the defect to lie at $x^- = X^-(y) = 0$ and $x^+ = X^+(y) = 0$. 

For the case of a flat defect, the operator $\Sigma_n^0$ breaks the conformal symmetry group down to $SO(2) \times SO(d-1,1)$, with the $SO(2)$ corresponding to the rotations of the plane orthogonal to the defect. This symmetry group suggests that a bulk dimension-$d$ CFT descends to a dimension $d-2$ defect CFT, which describes the excitations of the defect. We can thus use the language of boundary CFTs to analyze this problem. We will only give a cursory overview of this rich subject. For a more thorough review of the topic see \cite{Balakrishnan:2017aa, Bianchi_2016, Billo:2016aa}, and for additional background see \cite{Headrick:2010zt, Calabrese:2010he, Agon:2015ftl,Bousso:2014uxa}.  The important aspect for us will be the spectrum of operators that live on the defect.

The spectrum of operators associated to the twist defect was studied in \cite{Balakrishnan:2017aa}. In that work, techniques were laid out to understand how bulk primary operators induce operators on the defect. This can be quantitatively understood by examining the two-point function of bulk scalar operators in the limit that they both approach the defect. We imagine that as a bulk operator approaches the defect, we can expand in the transverse distance $|z|$ in a bulk to defect OPE so that
\begin{align}\label{eqn:defect expansion}
\lim_{|z| \to 0} \sum_{k=0}^{n-1} \mathcal{O}^{(k)}(z,\bz,y) \Sigma_n^0 = z^{-(\Delta_{\mathcal{O}} + \ell_{\mathcal{O}})}\bz^{-(\Delta_{\mathcal{O}}-\ell_{\mathcal{O}})}\sum_j C^j_{\mathcal{O}} z^{(\hat{\Delta}_j+\ell_j)/2}\bz^{(\hat{\Delta}_j - \ell_j)/2}\hat{\mathcal{O}}_j(y) \Sigma_n^0 
\end{align}
where $\Delta_{\mathcal{O}}$ is the dimension of the bulk operator, while $\hat{\Delta}_j$ is the dimension of the $j$th defect operator $\hat{\mathcal{O}}_j$. Every operator is also now labeled by its spin, $\ell$, under the $SO(2)$ rotations $z \to ze^{-i\phi}$. From the defect CFT point of view, the $SO(2)$ spin is an internal symmetry and the $\ell_j$'s are the defect operators' associated quantum numbers.  Notice that the $\mathbb{Z}_n$ symmetry has the effect of projecting out operators of non-integer spin. This is another reason for why the $\mathbb{Z}_n$ orbifolding is needed for treating the theory on the defect as a normal Euclidean CFT. 

Equation \eqref{eqn:defect expansion} suggests an easy way to obtain defect operators in terms of the bulk operators. Consider the lowest dimension defect operator $\hat{\Delta}_{\ell}$ of a fixed spin $\ell$. Then we can extract the defect operator via a residue projection, 
\begin{align}\label{eqn:bulktodefect}
\hat{\mathcal{O}_{\ell}}(0)\Sigma_n^0 = \lim_{|z|\rightarrow 0} \frac{|z|^{-\hat{\tau}_{\ell} + \tau_{\alpha}}}{2\pi i}\oint \frac{dz}{z}z^{-\ell + \ell_{\alpha}} \sum_{k=0}^{n-1}\mathcal{O}_{\alpha}^{(k)}(z,|z|^2/z,0)\Sigma_n^0 
\end{align} 
where $\hat{\tau}_{\ell}$ and $\tau_{\alpha}$ are the twists of the defect and bulk operators respectively. Note that these leading twist operators are necessarily defect primaries. 

Note that in general, due to the breaking of full conformal symmetry, $\hat{\Delta}_{\ell}$ will contain an anomalous dimension $\gamma_{\ell}(n)$. In this paper we will mainly be interested in the defect spectrum near $n = 1$ so after analytically continuing in $n$ we can expand $\gamma_{\ell}(n)$ around $n =1$ as $\gamma(n) = \gamma^{(0)}+\gamma^{(1)}(n-1) + \mathcal{O}((n-1)^2)$.
We now give a brief review of the various defect operators discovered in \cite{Balakrishnan:2017aa}.\footnote{See  \cite{Lemos:2017vnx} for a complementary method for computing the defect spectrum from the bootstrap and an appropriate Lorentzian inversion formula. It would be interesting to derive some of the results presented here in that language. } 

\subsection{Operators induced by bulk scalars or spin one primaries}
Associated to each bulk scalar $\phi$, or spin-one primary $V_{\mu}$, of dimension $\Delta_{\phi},\Delta_V$, the authors of \cite{Balakrishnan:2017aa} found a family of defect operators of dimension $\hat{\Delta}^{\ell}_{\phi, V} = \Delta_{\phi,V} -J_{\phi,V}+ \ell + \gamma^{(1)}_{\phi,V}(n-1) + \mathcal{O}((n-1)^2)$ with $SO(2)$ spin $\ell$ along with their defect descendants. Here $J_{\phi,V} = 0,1$ for $\phi$ and $V$ respectively and importantly $\ell \geq J$. The anomalous dimensions for the operators induced by bulk scalars, $\gamma_{\phi}$, are given in formula (3.25) of \cite{Balakrishnan:2017aa}. We will not be concerned with these two families in this paper.

\subsection{Operators induced by bulk primaries of spin $J\geq 2$}
For primary operators of spin $J\geq 2$, the authors of \cite{Balakrishnan:2017aa} again found a similar family of operators with dimensions $\hat{\Delta}_J^{\ell} = \Delta _J- J + \ell + \gamma^{(1)}_{J,\ell}  (n-1) + \mathcal{O}((n-1)^2)$ where $\ell \geq J$.

For a primary of spin $J\geq 2$, there are also $J-1$ ``new" operators with $SO(2)$ charge $J-1 \geq \ell \geq 1$. These ``displacement operators" can be written at integer $n$ as
\begin{align}\label{eqn:higherspind}
\hat{D}_{\ell}^{J} = i\oint d\bz \frac{\bz^{J-\ell-1}}{|z|^{\gamma_{J,\ell}(n)}} \sum_{k=0}^{n-1} \mathcal{J}_{+...+}^{(k)}(|z|^2/\bz,\bz)
\end{align}
where $J$ is the spin of the bulk primary $\mathcal{J}_{+...+}$ and $1 \leq \ell \leq J-1$ is the $SO(2)$ spin of the defect operator. The power of $|z|^{\gamma}$ accounts for the dependence of the defect operator dimension on $n$.

We will primarily be interested in the spectrum of $T_{++}$ on the defect for which there is only one displacement operator, $\hat{D}_+$. The displacement operator can also be equivalently defined in terms of the diffeomorphism Ward identity in the presence of the defect \cite{Billo:2016aa}
\begin{align}
\nabla^{\mu}\langle \Sigma_n^{\psi} T_{\mu\nu}\rangle = \delta(z,\bar{z})\langle \Sigma_n^{\psi} \hat{D}_{\nu}\rangle.
\end{align}
This implies that $\hat{D}_+$ corresponds to a null deformation of the orbifold partition function with respect to the entangling surface. In particular, entropy variations are given by $\hat{D}_+$ insertions in the limit $n \rightarrow 1$:
\begin{align}
\langle \Sigma_n^{\psi} \hat{D}_+(y)\rangle = (n-1) \langle \Sigma_n^{\psi} \rangle \frac{\delta S_{\psi}}{\delta x^+(y)} + \mathcal{O}((n-1)^2)
\end{align}   
The generalization to two derivatives is then just 
\begin{align}
\langle \Sigma^{\psi}_n \hat{D}_+(y) \hat{D}_+(y')\rangle = (n-1) \langle \Sigma_n^{\psi} \rangle \frac{\delta^2 S_{\psi}}{\delta X^+(y)X^+(y')} + \mathcal{O}((n-1)^2).
\end{align}   
We see importantly that statements about entropy variations can be related directly to displacement operator correlation functions.

\section{Towards saturation of the QNEC}\label{sec:saturation}
With the displacement operator in hand, we can now describe an argument for QNEC saturation. As just described, second derivatives of the entanglement entropy can be computed via two point functions of the defect CFT displacement operator. Thus, we are interested in proving the following identity:
\begin{align} \label{eqn:Main Result}
\lim_{n\rightarrow 1}\frac{1}{n-1}\langle \Sigma^{\psi}_n \hat{D}_+(y)\hat{D}_+(y')\rangle &= 2\pi \braket{\hat{T}_{++}(y)}_{\psi}\delta^{d-2}(y-y') \nonumber 
\\ &+ (\text{less divergent in }|y-y'|)
\end{align}
where $|\psi\rangle$ is any well-defined state in the CFT.

Since we are only interested in the short distance behavior of this equality - namely the delta function piece - we can examine the OPE of the displacement operators 
\begin{align}\label{eqn:dope}
\frac{1}{n-1}\hat{D}_+(y)\hat{D}_+(y') = \frac{1}{n-1}\sum_{\alpha} \frac{c_{\alpha}(n) \hat{\mathcal{O}}_{++}^{\alpha}(y)}{|y-y'|^{2(d-1)-\Delta_{\alpha} + \gamma_{\alpha}(n)}} + \text{descendants} 
\end{align}  
where $\Delta_{\alpha}$ is the dimension of the defect primary $\hat{\mathcal{O}}_{\alpha}$ at $n=1$ and $\gamma_{\alpha}(n)$ gives the $n$ dependence of the dimension away from $n=1$. We will refer to $\gamma_\alpha(n)$ as an anomalous dimension. Note that this is an OPE defined purely in the defect CFT. The $++$ labels denote the $SO(2)$ spin of the defect operator, which must match on both sides of the equation. The dimension of the displacement operators themselves are independent of $n$ and fixed by a Ward identity to be $d-1$.

At first glance, this equation would suggest that there are no delta functions in the OPE, only power law divergences. In computing the entanglement entropy, however, we are interested in the limit as $n \to 1$. In this limit, it is possible for a power law to turn into a delta function as follows: 
\begin{align} \label{eqn:deltaidentity}
\lim_{n\rightarrow 1} \frac{n-1}{|y-y'|^{d-2 - \gone(n-1)}} = \frac{S_{d-3}}{\gone} \delta^{(d-2)}(y-y')
\end{align}
where $\gamma = \gone (n-1) + \mathcal{O}((n-1)^2)$ and  $S_{d-3}$ in the area of the $d-3$ sphere. Comparison of equations \eqref{eqn:deltaidentity} and \eqref{eqn:dope} shows that a delta function can ``condense" in the $\hat{D}_+ \times \hat{D}_+$ OPE only if the OPE coefficient and anomalous dimension obey 
\begin{align}\label{eqn:ratio}
c_{\alpha}(n)/\gamma_{\alpha}(n) \sim (n-1) + \mathcal{O}((n-1)^2)
\end{align}
 as $n$ approaches $1$. 

This is, however, not sufficient for a delta function to appear in \eqref{eqn:dope} as $n \to 1$. We also need to have
\begin{align}
\Delta_{\alpha} = d \label{Relations}
\end{align}
at $n=1$. In other words, the defect operators we are looking for must limit to an operator of $SO(2)$ spin two and dimension $d$ as the defect disappears. Clearly, the $\ell =2 $ operator induced by the bulk stress tensor, $\hat{T}_{++}$, satisfies these conditions. Indeed, the first law of entanglement necessitates the appearance of $\hat{T}_{++}$ in the $\hat{D}_{+} \times \hat{D}_{+}$ OPE with a delta function (see Section \ref{firstlawsec} below). 

Our main claim, \eqref{eqn:Main Result}, is the statement that no other operator can show up in \eqref{eqn:dope} whose contribution becomes a delta function in the $n \to 1$ limit. In the rest of this section, we enumerate all the possible operators that could appear in the $\hat{D}_{+} \times \hat{D}_{+}$ OPE \eqref{eqn:dope}.

\subsection{Defect operators induced by low-dimension scalars}
If there exists a scalar operator of dimension $\Delta = d-2$, then the associated defect operator with $SO(2)$ spin $\ell = 2$ will have dimension $\Delta = d$ at leading order in $n-1$. This possibility was discussed in \cite{Leichenauer:2018obf}. The contribution of such an operator was found to drop out of the final quantity $\braket{T_{++}} - \frac{1}{2\pi} S_{++}''$ for holographic CFTs. We expect the same thing to happen in general CFTs in the presence of such an operator, so we ignore this possibility.

\subsection{$\ell =2$ operators induced by spin one primaries}
As discussed earlier, these defect operators have dimension $\hat{\Delta} = \Delta_{V} + 1 + \mathcal{O}(n-1)$. We see that for spin one primaries not saturating the unitarity bound, i.e. $\Delta_V >d-1$, these cannot contribute delta functions. Actually, since these operators exist in the CFT at $n=1$, we will argue in the next section that the first law of entanglement forces their OPE coefficients to be of order $(n-1)^2$. 

For spin-one primaries saturating the unitarity bound, $V_{\mu}$ is then the current associated to some internal symmetry. The entropy is uncharged under all symmetries, so such operators cannot contribute to $\hat{D}_+ \times \hat{D}_+$.

\subsection{$\ell = 2$ higher spin displacement operators}

The most natural candidate for contributions to the $\hat{D}_+ \times \hat{D}_+$ OPE are the $\ell = 2$ higher spin displacement operators discussed in the previous section. These operators are given by equation \eqref{eqn:higherspind}.

To show that such operators do not contribute delta functions to $\hat{D}_+ \times \hat{D}_+$, we need to argue that their dimensions $\Delta_n(\ell=2,J)$ do not limit to $d$ as $n\to 1$. As discussed in the previous section, the dimensions of the higher spin displacement operators are given by
\begin{align}\label{eqn:anomdim}
\Delta_n(\ell,J) = \Delta_J - J +\ell + \mathcal{O}(n-1).
\end{align}
The anomalous dimensions have not yet been computed but we expect them to be of order $n-1$, although we will not need this calculation here. The important point for us will be that in a CFT with a twist gap, the leading order dimension of these operators is 
\begin{align}
\Delta_n(2,J) = \tau_J +2 + \mathcal{O}(n-1) > d
\end{align}
assuming the twist of the bulk primaries satisfies $\tau_J > d-2$. Here we are using a result on the convexity of twist on the leading Regge trajectory for all $J$ proven in \cite{Costa_2017}. We see that the bulk higher spin operators would need to saturate the unitarity bound to contribute a delta function. Furthermore, there could be defect descendants of the form $(\partial_y^i \partial_y^i)^k \hat{D}^{J}_{++}(y)$. But such operators will necessarily contribute to the OPE with larger, positive powers of $|y-y'|$, hence they cannot produce delta functions.

\subsection{Nonlocal defect operators}\label{sec:nonstandard}

So far we have focused on the individual contribution of local defect operators and by power counting we see that these operators cannot appear in the diagonal QNEC. 
At fixed $n$, it is reasonable to conjecture that this list we just provided is complete. However we have not fully concluded that something more exotic does not appear in the OPE. As discussed in the introduction this possibility arises because the $n \rightarrow 1$ limit may not commute with the OPE. 

Indeed, we will find evidence that something non-standard does appear in the displacement OPE. In Section~\ref{sec:proof} and Section~\ref{sec:nearvacuum} we will present some computations of correlation functions of the displacement operator for particular states and entangling surfaces. In these specific cases we will be able to make the analytic continuation to $n \rightarrow 1$ before taking the OPE. In both cases, we find that the power laws as $ y_1 \rightarrow y_2$ are controlled by the dimensions associated to non-local spin-3 light ray operators \cite{Kravchuk:2018htv}. 
In the discussion section we will come back to the possibility that these contributions come from an infinite tower of the local defect operators that we have thus far enumerated. 
We conjecture that when this tower is appropriately re-summed, we will find these non-standard contributions to the entanglement entropy.

We will refer to these operators as \emph{nonlocal defect operators}, and we further conjecture that a complete list of such operators and dimensions is determined by the nonlocal $J=3$ lightray operators that appear in the lightray OPE of two averaged null energy operators as studied in \cite{Hofman_2008,Kologlu:2019mfz} for the CFT \emph{without} a defect. In order to give further evidence for this conjecture, in Section \ref{sec:proof} we will compute the analytic continuation of the spectrum of operators appearing around $n=1$ in the $\hat{D}_+ \times \hat{D}_+$ OPE by computing a fourth order shape variation of vacuum entanglement. Our answer is consistent with the above conjecture. While this relies on a specific continuation in $n$ (a specific choice of ``state dependence'') we think this is strong evidence that we have not missed anything.

 Before studying this nonlocal contribution further, we return to the local defect contribution where we would like to check that the ratio of $c(n)/\gamma(n)$ for $\hat{T}_{++}$ obeys \eqref{eqn:ratio}.

\section{Contribution of $\hat{T}_{++}$}\label{sec:Tpp}
In this section, we will review the first law argument which fixes the coefficient of the stress tensor defect operator to leading order in $n-1$. We will then use defect methods to demonstrate that the stress tensor does contribute with the correct ratio of $c(n)$ and $\gamma(n)$ to produce a delta function with the right coefficient demanded by the first law. To do this, we will make use of a slightly modified form of the usual diffeomorphism Ward identity in the presence of a twist defect that will compute $c(n)/\gamma(n)$. In Appendices \ref{cn} and \ref{gn}, we also explicitly calculate $c(n)$ and $\gamma(n)$ separately for the stress tensor and show that they agree with the result of this sub-section.

\subsection{The First Law}\label{firstlawsec}

A powerful guiding principle for constraining which defect operators can appear in the OPE \eqref{eqn:dope} is the first law of entanglement entropy. The entanglement entropy $S(\rho) = -\tr[\rho \log \rho]$, when viewed as the expectation value of the operator $-\log \rho$, is manifestly non-linear in the state. The first law of entanglement says that if one linearizes the von Neumann entropy about a reference density matrix - $\sigma$ - then the change in the entropy is just equal to the change in the expectation value of the vacuum modular Hamiltonian. Specifically it says that
\begin{align}\label{eqn:firstlaw}
\delta \tr[\rho \log \rho] = \tr[\delta \rho \log \sigma]
\end{align}
where $\rho = \sigma + \delta \rho$. 

The case we will be interested in here is when $\sigma$ is taken to be the vacuum density matrix for the Rindler wedge. The first law then tells us that the \emph{only} contributions to $\braket{\Sigma^{\psi}_n \hat{D}_+(y) \hat{D}_+(y')}$ that are linear in the state as $n \to 1$ must come from the shape variations of the vacuum modular Hamiltonian. 

The second shape derivative of the Rindler wedge modular Hamiltonian is easy to compute from the form of the vacuum modular Hamiltonian associated to generalized Rindler regions \cite{Wall:2011hj,Faulkner:2016mzt,Koeller:2017aa,Casini:2017aa}. Defining $\Delta \braket{H^{\sigma}_{\mathcal{R}}}_{\psi} = -\tr[\rho_{\mathcal{R}} \log \sigma_{\mathcal{R}}] + \tr[\sigma_{\mathcal{R}} \log \sigma_{\mathcal{R}}]$ to be the vacuum subtracted modular Hamiltonian for a general region $\mathcal{R}$ bounded by a cut of the $x^- = 0$ null plane, then we have the simple universal formula
\begin{align}
\frac{\delta^2 \Delta \braket{H^{\sigma}_{\mathcal{R}}}_{\psi}}{\delta X^{+}(y) \delta X^{+}(y')} = \frac{2\pi}{\hbar} \braket{T_{++}}_{\psi}\delta^{(d-2)}(y-y').
\end{align}

This is a simple but powerful constraint on the displacement operator OPE; it tells us that the only operator on the defect which is manifestly linear in the state as $n \to 1$ and appears in $\hat{D}_+ \times \hat{D}_+$ at $n=1$ is the stress tensor defect operator
\begin{align}\label{eqn:O++}
\hat{T}_{++} = \oint \frac{d\bz}{\bz |z|^{\gamma_n}} \sum_{j=0}^{n-1} T_{++}^{(j)}(|z|^2/\bz,\bz).
\end{align}
Thus, any other operator which appears in the OPE around $n=1$ must contribute in a manifestly non-linear fashion. Examining the list of local defect operators discussed in Section~\ref{sec:saturation} the only operators that are allowed by the above argument, aside from $\hat{T}_{++}$, are the higher spin displacement operators. As shown in \cite{Balakrishnan:2017aa} the limit $n \rightarrow 1$ of the expectation value of these operators give a contribution that is non-linear in the state. 

We will return to these state dependent operators in later sections. Now we check that indeed the stress tensor contributes with the correct coefficient.
 
\subsection{Using the modified Ward identity}
In Appendix \ref{modward}, we prove the following intuitive identity:
\begin{align}\label{Desired4} 
 \int d^{d-2}y' \langle \Sigma_n^0\hat{D}_+(y') \hat{D}_+(y) T_{--}(w,\bar{w}, 0)\rangle  &= -\partial_{\bar{w}}\langle \Sigma_n^0 \hat{D}_+(y) T_{--}(w,\bar{w},0)\rangle .
\end{align}

We now show that the identity \eqref{Desired4} allows us to compute the stress tensor contribution to the $\hat{D}_+ \times \hat{D}_+$ OPE, which can be written as: 
\begin{align}\label{eqn:Defect OPE}
\hat{D}_+(y) \hat{D}_+(y') \supset  \frac{c(n)}{|y-y'|^{d-2-\gamma(n)}}\hat{T}_{++}(y) + \ldots 
\end{align}
where we have focused on the $\hat{T}_{++}$ contribution and the ellipsis stand for the defect descendants of $\hat{T}_{++}$. We are free to ignore other defect primaries since they get projected out by the $T_{--}(w,\bw,0)$ insertion in \eqref{Desired4}.  Of course, since \eqref{Desired4} involves a $y$ integral, one might worry that we are using the OPE outside its radius of convergence. For now, we will follow through with this heuristic computation using the OPE. At the end of this subsection, we will say a few words about why this is justified.

Inserting \eqref{eqn:Defect OPE} into \eqref{Desired4} and ignoring the descendants, we find
\begin{align}
 \int d^{d-2}y'  \frac{c(n)}{|y-y'|^{d-2-\gamma(n)}}\braket{ \Sigma_n^0 \hat{T}_{++}(y) T_{--}(w,\bar{w}, 0)} = \frac{c(n)}{\gamma(n)} S_{d-3}\braket{ \Sigma_n^0 \hat{T}_{++}(y) T_{--}(w,\bar{w}, 0)} \label{Evaluate}
\end{align}
where $S_n$ is the area of the unit $n$-sphere. We can write $\hat{T}_{++}(y)$ in terms of $T_{++}$ integrated around the defect: 
\begin{align}
\hat{T}_{++}(y) = -\frac{1}{2\pi i}\sum_{k = 0}^{n-1}\oint \frac{d\bar{z}}{\bar{z}|z|^{\gamma(n)}}T^{(k)}_{++}(|z|^2/\bz, \bar{z}, y) \label{Wrapping}
\end{align}

We now take the $n \to 1$ limit of equation \eqref{Desired4}. Since the right hand side starts at order $(n-1)$, we see that $c(n)$ must begin at one higher order in $n-1$ than $\gamma(n)$. Generically we expect $\gamma(n)$ to begin at order $n-1$ and in Appendix \ref{gn} we will see that it does. We thus get the relation
\begin{align}
\frac{\ctwo}{\gone}\braket{\Sigma_1^0 \hat{T}_{++}(y)T_{--}(w,\bar{w},0)} =  -\partial_n \big\vert_{n=1} \partial_{\bar{w}}\braket{\Sigma_n^0 \hat{D}_+(y) T_{--}(w,\bar{w},0)}
\end{align}
where $c(n) = \cone (n-1) + \ctwo (n-1)^2 +...$ and $\gamma(n) = \gone (n-1) + ...$ .

At $n=1$, $\braket{ \Sigma_1^0\hat{T}_{++}(y)T_{--}(w,\bar{w},0)}$ is just the usual stress tensor 2-point function. Moreover, we can evaluate the right hand side of \eqref{Desired4} at order $(n-1)$ by following the steps leading up to eq. (3.31) of \cite{Balakrishnan:2017aa}. This leads to   
\begin{align}
\left. \partial_{\bar{w}}\langle \hat{D}_+(y) T_{--}(w,\bar{w},0)\rangle \right\vert_{|w|\rightarrow 0} &= i(n-1)\oint d\bar{z}\ \partial_{\bw} \left. \left(\int_{0}^{-\infty}\frac{d\lambda \ \lambda^2} {(\lambda-1)^2}\frac{c_T y^4}{4(w\bar{w} - w\bar{z}\lambda + y^2)^{d+2}}\right) \right \lvert_{|w|, |z| \to 0 } \nonumber
\\ &= -2\pi(n-1)\frac{c_T}{4}y^{-2d} 
\end{align}       

We are then left with the following expressions for $c_1$ and $c_2$:  
\begin{align}
\ctwo = \frac{2 \pi \gone}{S_{d-3}}, \ \cone = 0
\end{align}\label{c-over-g}
This is exactly what is needed in order to write \eqref{eqn:Defect OPE} near $y = y'$ as $\hat{D}_+(y) \hat{D}_+(y') \supset \delta^{(d-2)}(y-y') \hat{T}_{++}(y)$.

We now comment on the justification for using the $\hat{D}_+ \times \hat{D}_+$ OPE. Since the left hand side of \eqref{Desired4} involves a $y$ integral over the whole defect, one might worry that the we have to integrate outside the radius of convergence for the $\hat{D}_+ \times \hat{D}_+$ OPE. We see, however, that the $y$ integral produces an enhancement in $(n-1)$ only for the $T_{++}$ primary. In particular, this enhancement does not happen for the descendants of $T_{++}$. This suggests that if we were to plug in the explicit form of the defect-defect-bulk 3 point function into equation \eqref{Desired4} we would have seen that the $(n-1)$ enhancement comes from a region of the $y$ integral where $\hat{D}_{+}$ and $\hat{D}_{+}$ approach each other. We could then effectively cap the integral over $y$ so that it only runs over regions where the OPE is convergent and still land on the same answer. As a check of our reasoning, in Appendices \ref{cn} and \ref{gn}, we also compute the $c(n)$ and $\gamma(n)$ coefficients separately and check that they have the correct ratio.

%%%%%%%%%%%%%%%%%%%%%%%%%%%%%%%%%%%%%%%%%%%%

\section{Higher order variations of vacuum entanglement}\label{sec:proof}
In this section, we return to the possibility mentioned in Section \ref{sec:nonstandard} that something non-standard might appear in the displacement operator OPE. The authors of \cite{Balakrishnan:2017aa} argued that they had found a complete list of all local defect operators. This leaves open the possibility that the $n \to 1$ limit behaves in such a way that forces us to re-sum an infinite number of defect operators. 
In this Section and the next, we will find evidence that indeed this does occur. We will also give evidence that we have found a complete list of such nonlocal operators important for the $\hat{D}_+ \times \hat{D}_+$ OPE. In interacting theories with a twist gap this list does not include an operator with the correct dimension and spin that would contribute a delta function and violate saturation.

To get a better handle on what such a re-summed operator might be, we turn to explicitly computing the spectrum of operators in the $\hat{D} \times \hat{D}$ OPE.  To do this, we consider the defect four point function
\begin{align}
\mathcal{F}_n(y_1,y_2,y_3,y_4) = \braket{\Sigma_n^0 \hat{D}_+(y_1) \hat{D}_+(y_2) \hat{D}_-(y_3) \hat{D}_-(y_4)}.
\end{align}
We will consider configurations where $|y_1 - y_2| = |y_3 - y_4|$ are small but $|y_1 - y_4|$ is large. With these kinematics, we can use the $\hat{D} \times \hat{D}$ OPE twice and re-write the four point function as a sum over defect two point functions
\begin{align}\label{eqn:dpdpOPE}
\mathcal{F}_n = \sum_{\mathcal{O},\mathcal{O}'} \frac{c_{++}^{\mathcal{O}}(n) c_{--}^{\mathcal{O}'}(n)\braket{\Sigma_n^0\hat{\mathcal{O}}_{++}(y_2) \hat{\mathcal{O}}'_{--}(y_4)}}{|y_1 - y_2|^{2(d-1) + \hat{\Delta}_n^{\mathcal{O}}} |y_3-y_4|^{2(d-1) + \hat{\Delta}_n^{\mathcal{O}'}}}
\end{align}
where $\mathcal{O},\mathcal{O}'$ denote the local defect primaries and their descendants appearing in $\hat{D}\times \hat{D}$. We immediately see that by examining the powers of $|y_1 - y_2|$ appearing in $\mathcal{F}_n$, we can read off the spectrum of operators we are after. That is, at least before taking the limit $n \rightarrow 1$. 
We have not attempted to compute the OPE coefficients explicitly for all the local defect operators. This is left as an important open problem that would greatly clarify some of our discussion, but this is beyond the scope of this paper.

If we assume that the $n \rightarrow 1$ limit commutes with the OPE limit $y_1 \rightarrow y_2$ we can now find a contradiction. To see this contradiction, we can compute $\lim_{n \rightarrow 1} \mathcal{F}_n$ in an alternate manner holding $y_1,y_2$ fixed and compare to \eqref{eqn:dpdpOPE}. The main result we will find is that the divergences in $|y_1-y_2|$ appear to arise from defect operators of dimension $\Delta_{J_*} -J_*+2$ where $J_*=3$ and $\Delta_{J_*}$ is defined by analytically continuing the dimensions in \eqref{eqn:anomdim} to odd $J$ (recall that \eqref{eqn:anomdim} was only considered for even spins previously.) Generically we do not expect these particular dimensions to appear in the list of operator dimensions of the local defect operators that we enumerated. However we conjecture that by including such operator dimensions we complete the list of possible powers that can appear in the displacement OPE at $n=1$. 

This discussion further suggests that the final non-local defect operator that makes the leading contribution beside $T_{++}$ should be an analytic continuation in spin of the local higher spin displacement operators. We will come back to this possibility in the discussion. 

We now turn to computing $\mathcal{F}_n$ without using the defect OPE. In Appendix~\ref{app:calcFn}, we explicitly do the analytic continuation of $\mathcal{F}_n$, but here we simply state the answer. We find that $\mathcal{F}_n$ takes the form
\begin{align}\label{eqn:DDDD}
\mathcal{F}_n \sim (n-1) \int ds e^{-s}&\Braket{T_{--}(x^+=0,x^-=-1,y_3) \hat{\mathcal{E}}_+(y_1) \hat{\mathcal{E}}_+(y_2) T_{--}(x^+ = 0,x^- = -e^{-s},y_4)} \nonumber \\
&+ \mathcal{O}\left((n-1)^2\right), 
\end{align}
which can also be written as:
\begin{equation}
\mathcal{F}_n \sim (n-1) \frac{\left< \mathcal{E}_-(y_3) \hat{\mathcal{E}}_+(y_1) \hat{\mathcal{E}}_+(y_2) \mathcal{E}_-(y_4) \right>}{ {{\rm vol}\, SO(1,1)}}.
\end{equation}
The later division by the infinite volume of the 1 dimensional group of boosts is necessary to remove an infinity arising from an overall boost invariance of the four light-ray integrals. See for example \cite{Balakrishnan:2016ttg}. The un-hatted $\mathcal{E}_-$ operators represent half averaged null energy operators, integrated from the entangling surface to infinity.  Similar modifications to light-ray operators were used in \cite{Kologlu:2019mfz} in order to define their correlation functions and it is necessary here since otherwise the full light-ray operator would annihilate the vacuum. 

\begin{figure}[]
\center \includegraphics[width=.75\textwidth]{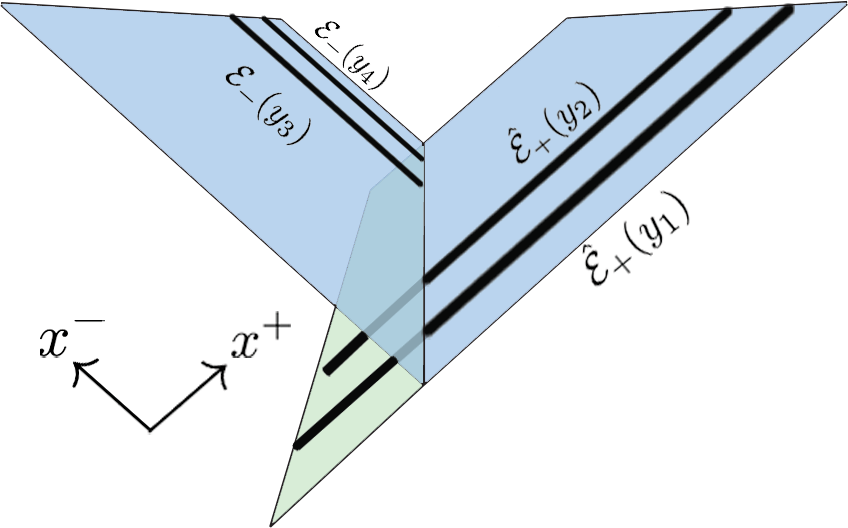}
\caption{The answer for the defect four point function $\mathcal{F}_n$ upon analytic continuation to $n=1$. We find that there are two insertions of half-averaged null energy operators, $\mathcal{E}_-$, as well as two insertions of $\hat{\mathcal{E}}_+$. Note that strictly speaking, in \eqref{eqn:DDDD}, the half-averaged null energy operators are inserted in the right Rindler wedge, but by CRT invariance of the vacuum, we can take the half-averaged null energy operators to lie in the left Rindler wedge instead, as in the figure.}
\label{fig2}
\end{figure}

We see that the effect of two $\hat{D}_+$ insertions was to create two $\hat{\mathcal{E}}_+$ insertions in the limit $n \rightarrow 1$. Thus considering the OPE of two displacement operators leads us to the OPE of two null energy operators. This object was studied in \cite{Hofman_2008} and more recently \cite{Kologlu:2019mfz}. These authors found that the two averaged null energy insertions can be effectively replaced by a sum over spin $3$ ``light-ray" operators, one for each Regge trajectory. In other words, 
\begin{align}
\hat{\mathcal{E}}_+(y_1) \hat{\mathcal{E}}_+(y_2) \sim \sum_{i} \frac{c_i \hat{\mathbb{O}}_i(y_2)}{|y_1 - y_2|^{2(d-2)-\tau_{\text{even},J=3}^i}}
\end{align}
where $\tau^i_{\text{even},J=3}$ is the twist of the even $J$ primaries on the $i$th Regge trajectory analytically continued down to $J=3$. A delta function can appear in this expression if $\tau_{\text{even},J=3}^i = d-2$, i.e. if the dimensions saturate the unitarity bound. 

Using the recent results in \cite{Costa_2017} again, we know that the twists on the leading Regge trajectory obey $\frac{d \tau(J)}{d J} \geq 0$ and $\frac{d^2 \tau(J)}{dJ^2} \leq 0$. Since the stress tensor saturates the unitarity bound, for a theory with a twist gap we know that $\tau_{\text{even},J=3}^i > d-2$, therefore there cannot be a delta function in $y_1 - y_2$. By the previous discussion then, formula \eqref{eqn:DDDD} suggests that there are no extra operators besides the stress tensor that produce a delta function. To give further evidence for this we next explicitly work out another case where we can compute the $n \rightarrow 1$ limit before we do the OPE and we find the same spectrum of operators. 

%%%%%%%%%%%%%%%%%%%%%%%%%%%%
\section{Near Vacuum States}\label{sec:nearvacuum}

We have just seen that the OPE of two displacement operators appears to be controlled by defect operators of dimension $\Delta_{J=3} -1$. As a check of this result, we will now independently compute the second variation of the entanglement entropy for a special class of states. In these states, we will again see the appearance of the OPE of two null energy operators $\hat{\mathcal{E}}_{+}(y) \hat{\mathcal{E}}_{+}(y')$. This again implies a lack of a delta function for theories with a twist gap. 

This computation is particularly illuminating in the case of free field theory where we can use the techniques of null quantization (see Appendix \ref{sec:null} for a brief review). Null quantization allows us to reduce a computation in a general state of a free theory to a near-vacuum computation. In this way we will also reproduce the computations in \cite{Bousso:2016aa} using a different method.

The state we will consider is a near vacuum state reduced to a right half-space
\begin{align}
\rho(\lambda) = \sigma + \lambda \delta \rho + \mathcal{O}(\lambda^2)
\end{align}
where $\sigma$ is the vacuum reduced to the right Rindler wedge. We can imagine $\rho(\lambda)$ as coming from the following pure state reduced to the right wedge 
\begin{align}\label{nearvacpurestate}
\ket{\psi(\lambda)} = \left( 1+ i\lambda \int dr d\theta d^{d-2} y g(r,\theta, y) \mathcal{O}(r, \theta, y) \right)\ket{\Omega} + \mathcal{O}(\lambda^2)
\end{align}
where $(r,\theta, y)$ are euclidean coordinates centered around the entangling surface and 
\begin{align}
\mathcal{O}(r, \theta, y) = \exp \left( i H^{\sigma}_{R} \theta  \right) \mathcal{O}(r, 0, y) \exp \left(-  i H^{\sigma}_{R} \theta  \right)
\end{align}
where $H_R^{\sigma}$ is the Rindler Hamiltonian for the right wedge.

From this expression for $\ket{\Psi(\lambda)}$, we have the formula
\begin{align}
\delta \rho = \sigma \int dr d\theta d^{d-2} y f(r,\theta,y) \mathcal{O}(r,\theta,y)
\end{align}
where 
\begin{align}
f(r,\theta,y) = i \left( g(r,\theta,y)-g(r,2\pi - \theta,y)^* \right) .
\end{align}
Note that $f$ obeys the reality condition $f(r,\theta,y) = f(r,2\pi - \theta,y)^*$.

We are interested in calculating the shape variations of the von-Neumann entropy. To this aim, since the vacuum has trivial shape variations we can compute the vacuum-subtracted entropy $\Delta S$ instead.  We start by using the following identity
\begin{align}
\Delta S = \tr \left( \left( \rho(\lambda)-\sigma \right) H^{\sigma} \right) - S_{\text{rel}}(\rho(\lambda)|\sigma).
\end{align}
We can now obtain $\Delta S$ to second order in $\lambda$. The vacuum modular Hamiltonian of the Rindler wedge is just the boost energy
\begin{align}
\tr \left[ (\rho(\lambda)-\sigma) H^{\sigma} \right] = \int d^{d-2}y \int dv v \tr \left[ \rho(\lambda) T_{++}(u=0,v,y) \right ]
\end{align}
where the computation of  $S_{\text{rel}}(\rho(\lambda)|\sigma)$ was done in Appendix B of \cite{Faulkner_2017}. There it was demonstrated that
\begin{align}\label{relativeentropy}
S_{\text{rel}}(\rho(\lambda)|\sigma) = -\frac{\lambda^{2}}{2} \int \frac{ds}{4 \sinh^{2}(\frac{s+i \epsilon}{2})} \tr \left[ \sigma^{-1} \delta \rho \sigma^{\frac{i s}{2\pi}} \delta \rho \sigma^{\frac{-i s}{2\pi}} \right] + \mathcal{O}(\lambda^3)
\end{align}

For a pure state like \eqref{nearvacpurestate}, we can instead write the above expression as a correlation function
\begin{align}\label{eqn:Srel}
S_{\text{rel}}(\rho|\sigma) = -\frac{\lambda^{2}}{2} \int d\mu \int \frac{ds}{4 \sinh^{2}(\frac{s+i\epsilon}{2})} \langle \mathcal{O}(r_{1},\theta_{1},y_{1}) e^{is \hat{K}}  \mathcal{O}(r_{2}, \theta_{2}, y_2) \rangle
\end{align}
where we have used the shorthand 
\begin{align}
\int d\mu = \int d r_{1,2} d\theta_{1,2} d^{d-2} y_{1,2} f(r_1,\theta_1,y_1)f(r_2,\theta_2,y_2)
\end{align}
and $\hat{K} = H^{\sigma}_{R} - H^{\sigma}_{L}$ is the full modular Hamiltonian associated to Rindler space.
This formula \eqref{eqn:Srel} and generalizations has been applied and tested in various contexts \cite{Faulkner:2014aa,Sarosi:2017rsq,Faulkner:2017tkh,Lashkari:2018tjh}. Most of these papers worked with perturbations about a state and a cut with associated to a modular Hamiltonian with a local flow such as the Rindler case. However it turns out that this formula can be applied more widely where $\hat{K}$ need not be local.\footnote{The only real subtlety is the angular ordering of the insertion of $\mathcal{O}$ in Euclidean. This can be dealt with via an appropriate insertion of the modular conjugation operator - a detail that does not affect the final result. We plan to work out these details in future work.}

We can thus safely replace the Rindler Hamiltonian in \eqref{eqn:Srel} with the Hamiltonian associated to an arbitrary cut of the null plane. This allows us to take shape deformations directly from \eqref{eqn:Srel}; by using the algebraic relation for arbitrary-cut modular Hamiltonians \cite{Casini:2017aa}
\begin{align}\label{eqn:algebra}
e^{-i \hat{K}(X^+)s} e^{i \hat{K}(0)s} = e^{i(e^s-1) \int dy \int d x^+ X^+(y) T_{++}(x^+)}
\end{align}
we have
\begin{align}\label{eqn:sreldp}
\frac{\delta^{2}S_{\text{rel}}(\rho | \sigma)}{\delta X^+(y) \delta X^+(y')} = \frac{\lambda^2}{2}\int d\mu \int ds e^{s} \langle \mathcal{O}(r_1, \theta_1, y_1) \mathcal{E}_{+}(y) \mathcal{E}_{+}(y') e^{is \hat{K}(X^+)}  \mathcal{O}(r_2, \theta_2, y_2) \rangle
\end{align}
where the states $\rho, \sigma$ depend implicitly on $X^+(y)$.\footnote{Note the similarity between \eqref{eqn:sreldp} and \eqref{eqn:DDDDfinal}. This is because one can view the defect four point function in \eqref{eqn:DDDD} as going to second order in a state-deformation created by stress tensors with a particular smearing profile.} Notice that upon taking the variations the double poles in the $1/\sinh^2(s/2)$ kernel of \eqref{relativeentropy} were precisely canceled by the factors of $e^s-1$ in the exponent of equation \eqref{eqn:algebra}.

\begin{figure}[]
\center \includegraphics[width=.75\textwidth]{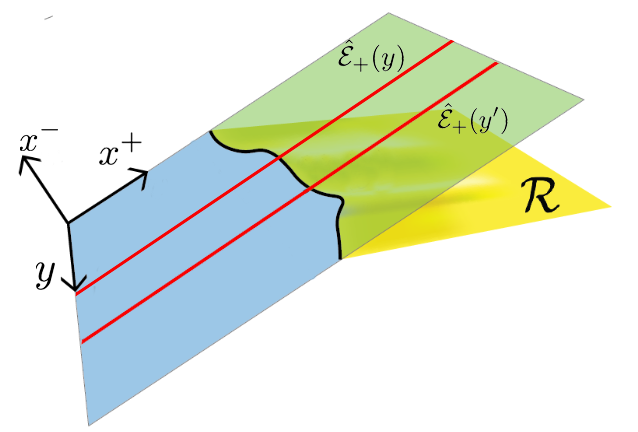}
\caption{For near vacuum states, the insertions of displacement operators limit to two insertions of the averaged null energy operators $\hat{\mathcal{E}}_+$.}
\label{fig1}
\end{figure}

This equation is the main result of this section. We see that taking shape derivatives of the entropy can for this class of states be accomplished by insertions of averaged null energy operators. This helps to explain the appearance and disappearance of extra delta functions as we change the coupling in a CFT continuously connected to a free theory. For example, in a free scalar theory, one can show that the OPE contains a delta function,
\begin{align}
\hat{\mathcal{E}}_{+}(y) \hat{\mathcal{E}}_{+}(y') \supset \delta^{d-2}(y-y').
\end{align}
This is consistent with the findings of \cite{Bousso:2015wca} where this extra delta function contribution to the QNEC was computed explicitly. To this aim, in Appendix \ref{sec:null}, we explicitly reproduce the answer in \cite{Bousso:2015wca} using the above techniques.

\section{Discussion}\label{sec:Discussion}

In this discussion, we briefly elaborate on the possible origin of the non-local operators whose dimensions we found in the displacement operator OPE considered in Sections \ref{sec:proof} and \ref{sec:nearvacuum}. As mentioned in the main text, the appearance of new operators is a bit puzzling since the authors in \cite{Balakrishnan:2017aa} found a complete set of defect operators as $n \rightarrow 1$. In other words, at fixed $n>1$, it should in principle be possible to expand these new operators as a (perhaps infinite) sum of $\ell =2$ defect operators. 

In particular, we expect them to be representable as an infinite sum over the higher spin displacement operators. We believe that it is necessary to do such an infinite sum before taking the $n \rightarrow 1$ limit, which entails that the OPE and replica limits do not commute.  This is why \cite{Balakrishnan:2017aa} did not find such operators. It also seems, given the non-trivial re-derivation of the results in \cite{Balakrishnan:2017aa} using algebraic tecniques in \cite{Ceyhan:2018zfg}, that these new non-local defect operators are not necessary for the limit $n \rightarrow 1$ limit of the bulk to defect OPE used in \cite{Balakrishnan:2017aa} to compute modular flow correlation functions. 

We give the following speculative picture for how the nonlocal defect operators might arise: 
\begin{align}
\label{highspin}
\hat{D}_+(y_1) \hat{D}_+(y_2) = \frac{c_{J=2}(n) \hat{T}_{++}}{|y_1-y_2|^{2(d-1)-\Delta_n^{J=2}}}+ \sum_{J=3}^{\infty} \frac{c_{J}(n) \hat{D}_{++}^{(J)}}{|y_1 - y_2|^{2(d-1) -\Delta_n^{J}}}
\end{align}
where we have suppressed the contribution of defect descendants. The latter sum in \eqref{highspin} comes from the spin $2$ displacement operators that come from the spin $J$ CFT operator. This is a natural infinite class of operators that one could try to re-sum should that prove necessary. 

In our calculations, we did not see any powers in $|y_1 - y_2|$ that could be associated to any individual higher spin displacement operator (as in the second term in \eqref{highspin}). Instead, in Section \ref{sec:proof} and Section \ref{sec:nearvacuum} after taking the $n \to 1$ limit we observed dimensions that did not belong to any of the known local defect operators. One possibility is that the higher spin operators in \eqref{highspin} re-sum into a new term that has a non-trivial interplay with the $n \to 1$ limit. One way this might happen is if the OPE coefficients of the higher spin displacement operators take the form
\begin{align}\label{eqn:scaling}
c_{J=2k}(n) \sim \frac{1}{(J-3)(n-1)^{J-3}}
\end{align}
so that they diverge as $n$ approaches $1$. Such a divergent expansion is highly reminiscent of the Regge limit for four point functions where instead the divergence appears from the choice of kinematics. This pattern of divergence where the degree increases linearly with spin can be handled using the Sommerfeld-Watson trick for re-summing the series. The basic idea is to re-write the sum as a contour integral in the complex $J$-plane. One then unwraps the contour and picks up various other features depending on the correlator. 

Our conjecture in \eqref{eqn:scaling} is that the other features which one encounters upon unwrapping the $J$ contour is quite simple: there is just one pole at $J=3$. Upon unwrapping the contour in the $J$-plane, we pick up the pole at $J=3$, which suggests that indeed these new divergences in $|y_1-y_2|$ are associated to operators which are analytic continuations in spin of the higher spin displacement operators. In this way we would reproduce the correct power law in $|y_1-y_2|$ as predicted for near vacuum states. 

Note that this needs to be true for \emph{any} CFT - not just at large $N$ or large coupling. The universality of this presumably comes from the universality of three point functions. Indeed, one can try to compute these OPE coefficients. We should consider the following three point function:
\begin{align}
\braket{\Sigma_n^0 \hat{D}_+(y_1) \hat{D}_+(y_2) \hat{D}_{--}^{(J)}(y_3)} \sim \frac{c_{J}(n) \braket{\Sigma_n^0 \hat{D}_{++}^{(J)}(y_2) \hat{D}_{--}^{(J)}(y_3)} }{|y_1-y_2|^{2(d-1)-\hat{\Delta}_n(J)}}
\end{align}

Via calculations based on the results in Appendix \ref{sec:multi-rep}, we find the three point function above in the the replica limit is: 
\begin{align}
\sim (n-1) \oint dw w^{J-3} \braket{\mathcal{J}_{-...-}(w,\bw=0,y_3) \hat{\mathcal{E}}_+(y_1) \mathcal{E}_+(y_2)} + \mathcal{O}((n-1)^2).
\end{align}
Naively, the full null energy operator $\hat{\mathcal{E}}_+(y_1)$ commutes with the half null energy operator $\mathcal{E}_+(y_2)$ and one can use the fact that $\hat{\mathcal{E}}_+(y_1) \ket{\Omega}=0$ to conclude that $c_{J}(n=1)$ vanishes. This seems to be incorrect however due to a divergence that arrises in the null energy integrals. Rather we claim that this coefficient diverges. The way to see this is to write
\begin{align}\label{eqn:JEE}
&\braket{\mathcal{J}_{-...-}(w,\bw=0,y_3) \hat{\mathcal{E}}_+(y_1) \mathcal{E}_+(y_2)} = \nonumber \\
& \int_{-\infty}^{\infty} dx^+_1 \int_0^{\infty} dx^+_2 \braket{\mathcal{J}_{-...-}(w,\bw=0,y_3) T_{++}(0,x_1^+,y_1) T_{++}(0,x_2^+,y_2)}.
\end{align}
We can now attempt to apply the bulk OPE between the two $T_{++}$'s which in these kinematics must become\footnote{To get the exact answer, one needs to account for all of the $SO(2)$ descendants in this OPE as well since they contribute equally to the higher spin displacement operator. We expect all of these descendants to have the same scaling behavior with $n-1$ and $J-3$.}
\begin{align}\label{eqn:TTOPE}
T_{++}(x^-=0,x^+_1,y_1)T_{++}(x^-=0,x^+_2,y_2)=\sum_{J=2}^{\infty} \frac{(x^+_{12})^{J-4} \mathcal{J}^{J}_{+...+}(x_2^+,y_2)}{|y_1 - y_2|^{2(d-1)-\hat{\Delta}_1(J)}} + \text{(descendants)}.
\end{align} 
where $\hat{\Delta}_1(J) = \Delta_J -J+2$. Plugging \eqref{eqn:TTOPE} into \eqref{eqn:JEE} and re-labeling $x_1 \to \lambda_1 x_2$, we see that for even $J\geq 3$, the $\lambda_1$ integral has an IR divergence 

One can cut-off the integral over $\lambda_1$ at some cutoff $\Lambda$. The answer will then diverge like
\begin{align}
& \frac{\left(\displaystyle{\int_{-\Lambda}^{\Lambda} d\lambda_1\, \lambda_1^{J-4}} \right)}{|y_1-y_2|^{2(d-1) - \hat{\Delta}_1(J)}} \times \int_0^{\infty} dx_2 x_2^{J-3} \braket{\mathcal{J}_{-...-}(w,\bw=0,y_3) \mathcal{J}_{+...+}(z=0,\bz=x_2^+,y_2)} \nonumber \\
&\sim \frac{\Lambda^{J-3}}{J-3} \int_0^{\infty} dx_2\, x_2^{J-3} \braket{\mathcal{J}_{-...-}(w,\bw=0,y_3) \mathcal{J}_{+...+}(z=0,\bz=x_2^+,y_2)} \times \frac{1}{|y_1-y_2|^{2(d-1) - \hat{\Delta}_1(J)}}.
\end{align}

The $\mathcal{J}-\mathcal{J}$ correlator on the right is precisely the order $n-1$ piece in $\braket{\Sigma_n^0 \hat{D}_{++}^{J} \hat{D}_{--}^{(J)}}$ so we find that the OPE coefficient scales like $c(n=1) \sim \frac{\Lambda^{J-3}}{J-3}$. 

Since $\Lambda$ is some auxiliary parameter, it is tempting to assign  $\Lambda \sim 1/(n-1)$; we then find the conjectured behavior in \eqref{eqn:scaling}. This is ad hoc and we do not have an argument for this assignmennt, except to say that the divergence is likely naturally regulated by working at fixed $n$ close to $1$. This is technically difficult  so we leave this calculation to future work.

\section*{Acknowledgments}
We would like to thank David Simmons-Duffin, Juan Maldacena, Sasha Zhiboedov, Aron Wall, and Stefan Leichenauer for discussions. We would also like to thank Stefan Leichenauer for helpful comments on the draft. The work of SB and TF was supported by the DOE contract SC0019183. The work of VC, AL, A.S-M. was supported in part by the Berkeley Center for Theoretical Physics; by the Department of Energy, Office of Science, Office of High Energy Physics under QuantISED Award DE-SC0019380 and contract DE-AC02-05CH11231; and by the National Science Foundation under grant PHY-1820912.
The work
of AL is also supported by the Department of Defense (DoD) through the National Defense
Science \& Engineering Graduate Fellowship (NDSEG) Program and the National Science Foundation under Grant No. NSF PHY-1748958.

\appendix 

\section{Modified Ward identity}\label{modward}
In this Appendix, we prove the following identity:
\begin{align}
\int d^{d-2}y' \langle \Sigma_n^0\hat{D}_+(y') \hat{D}_+(y) T_{--}(w,\bar{w}, 0)\rangle  &= -\partial_{\bar{w}}\langle \Sigma_n^0 \hat{D}_+(y) T_{--}(w,\bar{w},0)\rangle. \label{Desired} 
\end{align}
This is similar to the defect CFT ward identity of \cite{Billo:2016aa} except there is another insertion of the displacement operator. A priori it is not obvious that some form of the Ward identity carries through in the case where more than one operator is a defect operator. We will argue essentially that the second insertion of a $\hat{D}_+$ just comes along for the ride.

To show this, first we write the displacement operator as a stress tensor integrated around the defect: 
\begin{align}
\hat{D}_+(y) = i\oint d\bar{z} \  T_{++}(0,\bar{z}, y)
\end{align}
where we have suppressed the sum over replicas to avoid clutter. We will then argue that the following equality holds
\begin{align}\label{eqn:identity}
i&\lim_{\varepsilon \rightarrow 0}\oint_{\varepsilon > |\bar{z}|} d\bar{z} \int_{|y-y'|>\epsilon}d^{d-2}y' \langle \Sigma_n^0\hat{D}_+(y') T_{++}(0,\bar{z},y) T_{--}(w,\bar{w}, 0)\rangle \nonumber
\\ &= \int d^{d-2}y' \langle \Sigma_n^0 \hat{D}_+(y') \hat{D}_+(y) T_{--}(w,\bar{w}, 0)\rangle 
\end{align}  
for some appropriate $\varepsilon > 0$ that acts as the cutoff $|y'-y| > \varepsilon$. 

To see this, simply note that we can replace $T_{++}(0,\bar{z}, y)$ by a sum over local defect operators at $y$ using the bulk-defect OPE.  The important point is that this OPE converges because the $\bar{z}$ contour is always inside of the sphere of size $\varepsilon$ (by construction). We can take $|\bar{z}|$ to be arbitrarily small by making the size of the $\bar{z}$ contour as small as we like.  The $\bar{z}$ integral outside now simply projects the sum onto the displacement operator since we only consider the leading twist $d-2$ operators in the lightcone limit. Explicitly, we will be left with 
\begin{align}
i\lim_{\varepsilon \rightarrow 0} & \oint_{ \varepsilon > |\bar{z}|} d\bar{z} \int_{|y-y'|>\epsilon}d^{d-2}y' \langle \Sigma_n^0\hat{D}_+(y') T_{++}(0,\bar{z},y) T_{--}(w,\bar{w}, 0)\rangle \nonumber
\\ &= \lim_{\epsilon \to 0}\int_{|y-y'|>\epsilon} d^{d-2}y' \langle \Sigma_n^0 \hat{D}_+(y') \hat{D}_+(y) T_{--}(w,\bar{w}, 0)\rangle .
\end{align}        
Note that perturbatively around $n=1$, the integral over $|y-y'|>\epsilon$ will miss the delta function contribution to the $\hat{D}_+ \times \hat{D}_+$ OPE. Non-perturbatively away from $n=1$, however, there are no delta-function singularities in $|y-y'|$ present in the $\hat{D}_+ \times \hat{D}_+$ OPE. In what follows, we must be careful to take $\epsilon \to 0$ \emph{before} taking $n \to 1$. 

Using this identity, we can view the displacement-displacement-bulk three point function as the contour integral of a displacement-bulk-bulk three point function. We can then use the regular displacement operator Ward identity on the latter three point function. This Ward identity follows from general diffeomorphism invariance \cite{Billo:2016aa}. To do this, define the deformation vector field 
\begin{align}
\xi(y') = f(y')\partial_+\text{ with }f(y') = \Theta(|y'-y| -\varepsilon).
\end{align} For this deformation, the Ward identity takes the form 
\begin{align}
i&\oint_{\varepsilon > |\bar{z}|} d\bar{z} \int_{|y-y'|>\epsilon}d^{d-2}y' \langle \Sigma_n^0\hat{D}_+(y') T_{++}(0,\bar{z},y) T_{--}(w,\bar{w}, 0)\rangle  \nonumber
 \\ &= -f(0)\partial_{\bar{w}}\langle \Sigma_n^0\hat{D}_+(y) T_{--}(w,\bar{w},0)\rangle - i\oint d\bar{z} f(y)\partial_{\bar{z}}\langle \Sigma_n^0T_{++}(0,\bar{z},y) T_{--}(w,\bar{w}, 0)\rangle \nonumber
 \\ & -i\int_{\mathcal{M}_n}d^d x' \oint d\bar{z}\ \langle T_{++}(0,\bar{z},y) T_{--}(w,\bar{w}, 0) T^{\mu\nu}(x')\partial_{\mu}\xi_{\nu}(x') \rangle \label{Stuff}
\end{align}
where $\mathcal{M}_n$ is the full replica manifold. 

The second term on the right hand side of the equality vanishes because $f(y) = 0$. Since $f(0) = 1$ by construction we just need to argue that the last term in \eqref{Stuff} vanishes. 

\subsubsection*{Arguing the last term vanishes}
It is tempting at this stage to integrate by parts on the last term and conclude that this vanishes as one sends $\varepsilon \to 0$. Unfortunately, the last term in \eqref{Stuff} can produce $1/\varepsilon$ enhancements due to $T_{i+}$ operator coming $\varepsilon$ close to $T_{++}$. Therefore one must take care to first do the $x'$ integral and then take the $\varepsilon \rightarrow 0$ limit when evaluating this term. 

To do so, note that   
\begin{align}
T^{\mu\nu}(x')\partial_{\mu}\xi_{\nu}(x') &= \frac{1}{2}T_{i+}(x')\hat{n}^i\delta(|y'-y| - \varepsilon)
 \label{Diffeo Term}
\end{align}
where $\hat{n}^i = (y'-y)^i/|y'-y|$.
We then have the following
\begin{align}
&\int_{\mathcal{M}_n}d^d x' \oint d\bar{z}\ \langle T_{++}(0,\bar{z},y) T_{--}(w,\bar{w}, 0) T^{\mu\nu}(x')\partial_{\mu}\xi_{\nu}(x') \rangle \nonumber
\\ &= \frac{1}{2}\varepsilon^{d-3}\int \rho' d\rho' d\theta' \oint d\bar{z}\int d^{d-3}\vartheta' \ \hat{n}^i \langle T_{++}(0,\bar{z},y) T_{--}(w,\bar{w}, 0)T_{i+}(|\vec{y} + \vec{\varepsilon}|, \vartheta'_{\vec{\varepsilon}}, \rho' e^{-i\theta'}, \rho' e^{-i\theta'})  \rangle  \label{Stuffs}
\end{align}       
where $|\vec{\varepsilon}| = \varepsilon$. In going to the second line we have done the coordinate transformation $x'^+ = \rho' e^{-i\theta'}, \ x'^{-} = \rho' e^{i\theta'}$ because we are in the Euclidean section, and in going to the last line we have written $y'$ in spherical coordinates on the defect. At this point we can safely send $w,\bar{w}\rightarrow 0$ so that $T_{--}$ is simply fixed at the origin. Then, in particular, let us focus on 
\begin{align}
\int d\theta'  \oint d\bar{z}\  \langle T_{++}(0,\bar{z},y) T_{--}(0)T_{i+}(|\vec{y} + \vec{\varepsilon}|, \vartheta'_{\vec{\varepsilon}}, \rho' e^{-i\theta'}, \rho' e^{-i\theta'})  \rangle.
\end{align}  
It is easy to see that this identically vanishes from the boost weights of the quantities involved. Specifically, $T_{++}$ will yield a factor of $e^{2i\theta'}$, $T_{i+}$ will yield a factor of $e^{i\theta'}$, $T_{--}$ does not have a boost weight since it is fixed at the origin, and the measure $d\bar{z}$ will yield a factor of $e^{-i\theta'}$ so overall we will have $\int_0^{2\pi} d\theta' e^{i\theta'} = 0$. Therefore \eqref{Stuffs} is zero for any $\varepsilon$.     

Thus, the identity in \eqref{Stuff} becomes
\begin{align}
&i\lim_{\epsilon \to 0}\oint_{\varepsilon > |\bar{z}|} d\bar{z} \int_{|y-y'|>\epsilon}d^{d-2}y' \langle \Sigma_n^0\hat{D}_+(y') T_{++}(0,\bar{z},y) T_{--}(w,\bar{w}, 0)\rangle \nonumber \\
&  = -\partial_{\bar{w}}\langle \Sigma_n^0 \hat{D}_+(y) T_{--}(w,\bar{w},0)\rangle
\end{align}
which, using \eqref{eqn:identity}, proves \eqref{Desired}.

%%%%%%%%%%%%%%%%%%%%%%%%%%%%%%%%%%%%%%%%%%
\section{Analytic Continuation of a Replica Three Point Function}\label{sec:multi-rep}
In this section, we analytically continue a general $\mathbb{Z}_n$-symmetrized three point function of the form\footnote{Note that we are writing this as a thermal three point function on $\mathbb{H}_{d-1} \times S_1$, which is related to the flat space replica answer via conformal transformation. For a review of the relevant conformal factors, which we suppress for convenience, see \cite{Faulkner:2014aa}.}
\begin{align}\label{eqn:threepoint}
\mathcal{A}^{(3)}_n = n\sum_{j=0}^{n-1}\sum_{k=0}^{n-1} \tr \left[ e^{-2\pi n H} \mathcal{T} \mathcal{O}_a(0)\mathcal{O}_b(\tau_{ba}+2\pi j)\mathcal{O}_c(\tau_{ca}+2\pi k) \right] 
\end{align}
where $H$ is the vacuum modular Hamiltonian for the Rindler wedge and $\mathcal{T}$ denotes Euclidean time ordering with respect to this Hamiltonian. 

Following \cite{Faulkner:2014aa}, we begin by rewriting the the $j$-sum as as a contour integral
\begin{align}\label{eqn:wrapped}
\frac{n}{2\pi i}\sum_{k=0}^{n-1} \oint_{C_b} ds_b \frac{\tr \left[ e^{-2\pi n H} \mathcal{T} \mathcal{O}_a(0)\mathcal{O}_b(-is_b)\mathcal{O}_c(2\pi k + \tau_{ca}) \right]}{(e^{s_b - i\tau_{ba}}-1)}
\end{align}
where the contour $C_b$ wraps the $n$ poles at $s_b =  i(2\pi j + \tau_{ba})$ for $j = 0, ..., n-1$. We will now unwrap the $s_b$ contour integral in the complex plane, but will need to be careful as the analytic structure of the integrand in \eqref{eqn:wrapped} is non-trivial as a function of $s_b$; the integrand has poles at $s_b = i(2\pi j + \tau_{ba})$ and light-cone branch cuts lying along the lines $\Im s_b = 0, 2\pi n$ and $\Im s_b = 2\pi k + \tau_{ca}$ for a fixed $k$. The first two branch cuts were discussed in \cite{Faulkner:2014aa}. The third (middle in the figure) branch cut arises from singularities due to $\mathcal{O}_b$ and $\mathcal{O}_c$ lying on the same light-cone.

\begin{figure}
	\centering
	\includegraphics[scale=.9]{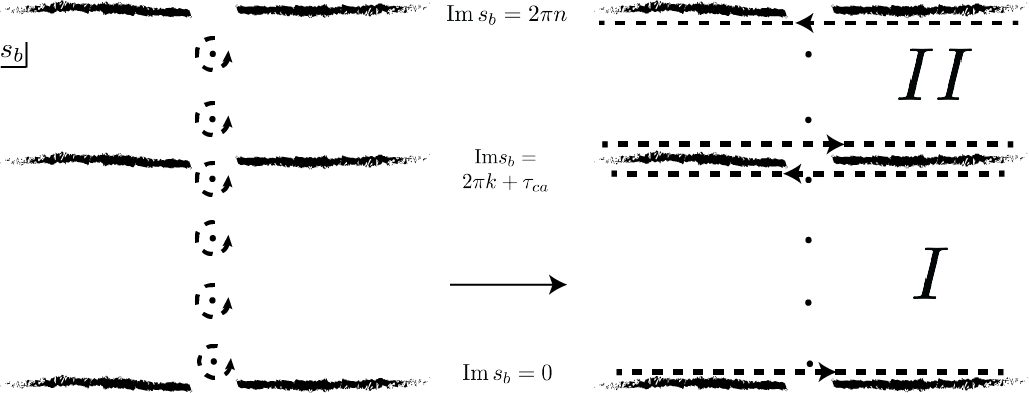}
	\caption{The analytic structure of the integral in equation \eqref{eqn:wrapped} represented in the $s_b$ plane for fixed $s_k = i(2\pi k + \tau_{ca})$ for $n=6$. The dots represent poles at $s_b = i(2\pi j + \tau_{ba})$ and the fuzzy lines denote light-cone branch cuts. The bottom and top branch cuts (which are identified by the KMS condition) arise from $\mathcal{O}_b$ becoming null separated from $\mathcal{O}_a$ and the middle branch cut arises from $\mathcal{O}_b$ becoming null separated from $\mathcal{O}_c$. Note that in this figure, $k = 3$ and $\tau_{ca}>\tau_{ba}>0$. We start with the contour $C_b$ represented by the dashed lines encircling the poles at $s_b = i (2\pi j + \tau_{ba})$ and unwrap so that it just picks up contributions from the branch-cuts. Region $I$ corresponds to the ordering $\mathcal{O}_a\mathcal{O}_b\mathcal{O}_c$ whereas region $II$ corresponds to $\mathcal{O}_a\mathcal{O}_c\mathcal{O}_b$.}.
	\label{fig:cbcontour}
\end{figure}

We can unwrap the $C_b$ contour now so that it hugs the branch cuts as in the right-hand panel of Figure \ref{fig:cbcontour}. We will then be left with a sum of four Lorentzian integrals 
\begin{align}\label{eqn:unwrapped}
& \frac{n}{2\pi i} \sum_{k=0}^{n-1} \tr \left [  e^{-2\pi n H} \int_{-\infty}^{\infty} ds_b \times \right. \nonumber \\
& \frac{\mathcal{O}_a(0)\mathcal{O}_b(-is_b+\epsilon_j)\mathcal{O}_c(2\pi k + \tau_{ca})}{(e^{s_b - i\tau_{ba}}-1)}-\frac{\mathcal{O}_a(0)\mathcal{O}_b(-is_b+2\pi i k + \tau_{ca}-\epsilon)\mathcal{O}_c(2\pi k + \tau_{ca})}{(e^{s_b+2\pi i k + \tau_{ca}-i\epsilon- i\tau_{ba}}-1)} \nonumber \\
& \left. +\frac{\mathcal{O}_a(0)\mathcal{O}_c(2\pi k + \tau_{ca})\mathcal{O}_b(-is_b + 2\pi k + \tau_{ca}+\epsilon)}{(e^{s_b+2\pi i k + \tau_{ca} +i\epsilon - i\tau_{ba}}-1)} - \frac{\mathcal{O}_a(0)\mathcal{O}_c(2\pi k + \tau_{ca})\mathcal{O}_b(-is_b+ 2\pi n-\epsilon)}{(e^{s_b+i2\pi n-i\epsilon - i\tau_{ba}}-1)}\right],
\end{align}
where we have set $2\pi k + \tau_{ca} = -is_c$ since the $C_c$ contour still wraps the poles at these values.

We now need to make a choice about how to do the analytic continuation in $n$. The usual prescription, which was advocated for in \cite{Faulkner:2014aa}, is to set $e^{2\pi i n}=1$ in the last term of \eqref{eqn:unwrapped}. We will follow this but also make one other choice. In the second and third terms in the integrand of \eqref{eqn:unwrapped} we make the choice to set $e^{2\pi i k} =1$ for all $k = 0, ..., n-1$.

Making this analytic continuation, we can now re-write the $k$-sum as a contour integral over $s_c$ along some contour $C_c$. Unwrapping this $s_c$ contour into the Lorentzian section, and after repeated use of the KMS condition to push operators back around the trace, we land on the relatively simple formula
\begin{align}\label{eqn:3pointfinal}
&\mathcal{A}_n^{(3)} = \nonumber \\
&\frac{-n}{4\pi^2}  \int_{-\infty}^{\infty} ds_c ds_b \tr \left[ e^{-2\pi n H}  \left(\frac{[[\mathcal{O}_a(0),\mathcal{O}_b(-is_b)],\mathcal{O}_c(-is_c)]}{(e^{s_b - i\tau_{ba}}-1)(e^{s_c - i\tau_{ca}}-1)} - \frac{[\mathcal{O}_a(0),[\mathcal{O}_b(-is_b-is_c),\mathcal{O}_c(-is_c)]]}{(e^{s_b+i\tau_{ca} - i\tau_{ba}}-1)(e^{s_c - i\tau_{ca}}-1)}\right) \right]
\end{align}
In deriving this formula, we have assumed $\tau_{ba}>0$ and $\tau_{ca}>0$ but we have not yet assumed any relationship between $\tau_{ba}$ and $\tau_{ca}$. This formula is the full answer. One could stop here, but we will massage this formula into a slightly different form for future convenience. Instead of following \cite{Faulkner:2014aa} and applying $\partial_n$ at this stage, which drops down powers of $H$, we will use a slightly different (although equivalent) technique.

We first focus on re-writing the two Lorentzian integrals in region $I$ of Figure \ref{fig:cbcontour} as one double integral. 

\subsubsection*{Region $I$}
Before re-writing the $k$-sum as a contour integral, the integrals in region $I$ are\footnote{For ease of notation, we have switched to $\braket{\mathcal{O}_1 \mathcal{O}_2 \mathcal{O}_3}_n = \tr [ e^{-2\pi n H} \mathcal{O}_1 \mathcal{O}_2 \mathcal{O}_3]$.}
\begin{align}\label{eqn:unwrapped1}
&\frac{n}{2\pi i} \sum_{k=0}^{n-1} \int_{-\infty}^{\infty} ds_b \left( \frac{\braket{\mathcal{O}_a(0)\mathcal{O}_b(-is_b)\mathcal{O}_c(2\pi k +\tau_{ca})}_n}{(e^{s_b - i\tau_{ba}}-1)}-\frac{\braket{\mathcal{O}_a(0)\mathcal{O}_b(-is_b+ 2\pi k + \tau_{ca} - \epsilon)\mathcal{O}_c(2\pi k + \tau_{ca})}_n}{(e^{s_b+ i\tau_{ca}- i\tau_{ba}}-1)} \right)
\end{align}
where as before we have set $e^{2\pi i k} = 1$ in the second term. The goal will be to make the denominators in these two terms the same so that we may combine their numerators. We will try to shift the $s_b$ contour in the second term by an amount $-i \tau_{ca}$, making sure not to cross any poles or branch cuts. To make our lives easier, we will assume a fixed ordering of the operators. For now, we will pick $\tau_{ca} > \tau_{ba}>0$. Note that any other ordering can be reached just by exchanging the $a,b,c$ labels.

\begin{figure}
	\centering
	\includegraphics[scale=.9]{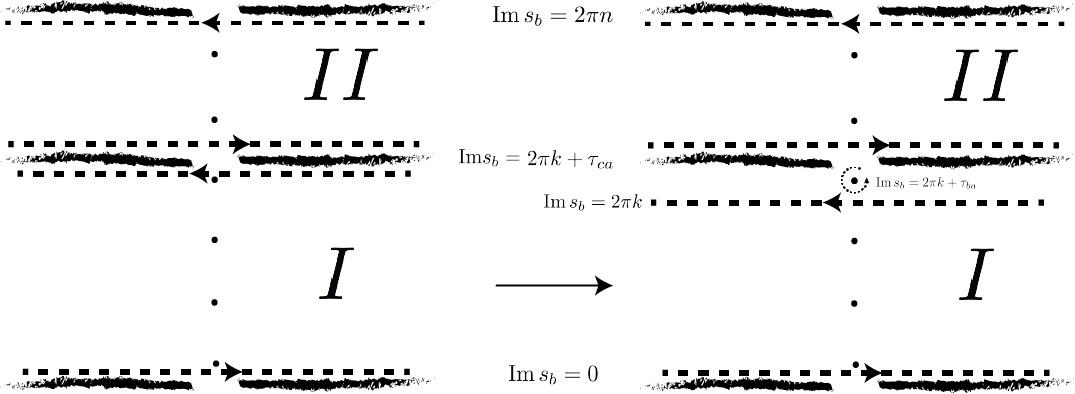}
	\caption{This figure illustrates the contour shift $s_b \to s_b -i\tau_{ca}$ done at the cost of picking up the pole at $s = i(2\pi k +\tau_{ba})$ when $\tau_{cb} = \tau_{ca}- \tau_{ba}>0$.}
	\label{fig:contour2}
\end{figure}

In this ordering, sending $s_b \to s_b - i\tau_{ca}$ crosses a pole at $\Im s_b = 2\pi k + \tau_{ba}$.  This contour shift is illustrated in Figure \ref{fig:contour2}.
After doing this shift, we get 
\begin{align}\label{eqn:unwrapped2}
&\frac{n}{2\pi i}\sum_{k=0}^{n-1}\int_{-\infty}^{\infty} ds_b \left( \frac{\braket{\mathcal{O}_a(0)\mathcal{O}_b(-is_b)\mathcal{O}_c(2\pi k + \tau_{ca})}_n - \braket{\mathcal{O}_a(0)\mathcal{O}_b(-is_b + 2\pi k)\mathcal{O}_c(2\pi k + \tau_{ca})}_n}{(e^{s_b - i\tau_{ba}}-1)} \right)\nonumber \\
& + \theta(\tau_{cb})\times (\text{terms with } j=k).
\end{align}
where we will mostly neglect the extra term coming from picking up the pole since it will not be important for most calculations we are interested in. We will refer to these terms as the ``replica diagonal terms" since they arise from terms in the double sum over $j,k$ in \eqref{eqn:threepoint} where $j = k$.

The numerator for the first term in equation \eqref{eqn:unwrapped2} then looks like the integral of a total derivative in some auxiliary parameter $t_b$ which we write as
\begin{align}\label{eqn:tb}
&\frac{-n}{2\pi i}\sum_{k=0}^{n-1} \int_{-\infty}^{\infty} ds_b \int_0^{i 2\pi k} dt_b \left( \frac{\frac{d}{dt_b} \braket{\mathcal{O}_a(0)\mathcal{O}_b(-is_b-it_b)\mathcal{O}_c(2\pi k+\tau_{ca})}_n}{(e^{s_b - i\tau_{ba}}-1)} \right).
\end{align}
Since $t_b$ shows up on equal footing with $s_b$ in the numerator, we see we can re-write the derivative in $t_b$ as one in $s_b$. Integrating by parts and dropping the boundary terms\footnote{We will drop boundary terms at large Lorentzian time everywhere throughout this discussion, as we expect thermal correlators to fall off sufficiently quickly \cite{Faulkner:2014aa}.}, we get 

\begin{align}\label{eqn:tb}
&\frac{-n}{2\pi i}\sum_{k=0}^{n-1}\int_{-\infty}^{\infty} ds_b \int_0^{i 2\pi k} dt_b \frac{\braket{\mathcal{O}_a(0)\mathcal{O}_b(-is_b-it_b)\mathcal{O}_c(2\pi k+\tau_{ca})}_n}{4\sinh^2((s_b - i\tau_{ba})/2)}.
\end{align}

We are now ready, as above, to turn the sum over $k$ into a contour integral over some Lorentzian parameter $s_c$. We can then execute the same trick as before: we re-write two terms as the boundary terms of one integral in some new auxiliary parameter $t_c$. After all of this, the answer we find is the relatively simple result for region $I$
\begin{align}\label{eqn:regioni}
\text{region I} = &\frac{-n}{4\pi^2}\int_{-\infty}^{\infty} ds_c ds_b \int_{0}^{i 2\pi (n-1)}dt_c \int_0^{s_c+t_c} dt_b \frac{\braket{\mathcal{O}_a(0)\mathcal{O}_b(-is_b-it_b)\mathcal{O}_c(-is_c-it_c+\tau_{ca})}_n}{16\sinh^2((s_b - i\tau_{ba})/2)\sinh^2((s_c - i\epsilon)/2)} \nonumber \\
& + \theta(\tau_{cb}) \times (\text{terms with } j=k).
\end{align}
Note that the quadruple integral term is manifestly order $n-1$ because of the limits on the $t_c$ integral.

\subsubsection*{Region $II$}
In region $II$ of Figure \ref{fig:cbcontour}, the calculations are exactly analogous, except now the ordering of the operators is different. We find that (up to terms that again come from picking up specific poles) the answer for region $II$ is
\begin{align}\label{eqn:regionii}
&\text{region II} =\nonumber \\
&\frac{-n}{4\pi^2}\int_{-\infty}^{\infty} ds_c ds_b \int_{0}^{i 2\pi (n-1)}dt_c \int_{s_c+t_c+i2\pi}^{i2\pi n} dt_b \frac{\braket{\mathcal{O}_a(0)\mathcal{O}_c(-is_c-it_c+\tau_{ca})\mathcal{O}_b(-is_b-it_b)}_n}{16\sinh^2((s_b - i\tau_{ba})/2)\sinh^2((s_c - i\epsilon)/2)} \nonumber \\
& + \theta(\tau_{bc}) \times (\text{terms with } j=k).
\end{align}

\subsubsection*{Combining Regions I and II}
Adding the Region I and Region II contributions, we get for the non-replica diagonal contributions to $\mathcal{A}_n^{(3)}$
\begin{align}
	&\frac{n}{4\pi^2}\int_{-\infty}^{\infty} ds_c ds_b \int_{0}^{i 2\pi (n-1)}dt_c \int_{0}^{s_c+t_c} dt_b \frac{\braket{[\mathcal{O}_b(-is_b-it_b),\mathcal{O}_a(0)]\mathcal{O}_c(-is_c-it_c+\tau_{ca})}_n}{16\sinh^2((s_b - i\tau_{ba})/2)\sinh^2((s_c - i\epsilon)/2)}  \nonumber \\
	 + &\frac{n}{4\pi^2}\int_{-\infty}^{\infty} ds_c ds_b \int_{0}^{i 2\pi (n-1)}dt_c \int^{s_c+t_c+i2\pi(1-n)}_{s_c+t_c} dt_b \frac{\braket{\mathcal{O}_b(-is_b-it_b)\mathcal{O}_a(0)\mathcal{O}_c(-is_c-it_c+\tau_{ca})}_n}{16\sinh^2((s_b - i\tau_{ba})/2)\sinh^2((s_c - i\epsilon)/2)} 
	 \label{eqn:nondiag}
\end{align}
where we used the KMS condition to push $\mathcal{O}_b$ around to the left of $\mathcal{O}_a$ in \eqref{eqn:regionii}. We then split the $t_b$ contour in \eqref{eqn:regionii} into two pieces, one purely Lorentzian integral from $t_b = 0$ to $t_b = s_c +t_c$ and another purely Euclidean integral from $t_b = s_c+t_c$ to $t_b = s_c+t_c + 2\pi i (n-1)$. Again, this is the full answer for the replica three point function, $\mathcal{A}_n^{(3)}$, at all $n$ excluding the replica diagonal terms.

From this we can compute the leading order in $n$ correction to the three-point function (dropping the diagonal terms). Taking an $n$-derivative and setting $n \to 1$, the total correction is 
\begin{align}\label{eqn:correction}
&\mathcal{A}_n^{(3)} \sim \frac{i(n-1)}{2\pi}\int_{-\infty}^{\infty} ds_c ds_b \int_0^{s_c} dt_b \frac{\braket{[\mathcal{O}_b(-is_b-it_b),\mathcal{O}_a(0)]\mathcal{O}_c(-is_c+\tau_{ca})}_1}{16\sinh^2((s_b - i\tau_{ba})/2)\sinh^2((s_c - i\epsilon)/2)}  \nonumber \\
& + \text{(replica diagonal terms)} + \mathcal{O}\left((n-1)^2\right).
\end{align}

\subsubsection*{Replica Diagonal Terms}
For future reference, we now list the replica diagonal (or $j=k$) terms that we have suppressed. In the order we considered above, we have
\begin{align}
&n\theta(\tau_{cb})\theta(\tau_{ba})\sum_{k=0}^{n-1} \braket{\mathcal{O}_a(0) \mathcal{O}_b(2\pi k + \tau_{ba}) \mathcal{O}_c(2\pi k + \tau_{ca})}_n \nonumber \\
& = n\theta(\tau_{cb})\theta(\tau_{ba}) \bigg( \braket{\mathcal{O}_a(0)\mathcal{O}_b(\tau_{ba})\mathcal{O}_c(\tau_{ca})}_n - \nonumber \\
& \left. \frac{1}{2\pi i}\int_{i2\pi}^{i2\pi n} dt_c \int_{-\infty}^{\infty} ds_c \frac{\braket{\mathcal{O}_a(0)\mathcal{O}_b(-is_c-it_c-\tau_{cb})\mathcal{O}_c(-is_c-it_c)}_n}{4\sinh^2((s_c - i\tau_{ca})/2)}  \right).
\end{align}
Again, other orderings can be found just by swapping the $a,b,c$ labels accordingly. Note that at $n=1$, the integral term vanishes and the answer reduces to the angular ordered three-point function as expected.

\section{Explicit Calculation of $c^{(2)}$}\label{cn}
In this section, we compute the OPE coefficient of $\hat{T}_{++}$ in the $\hat{D}_+ \times \hat{D}_+$ OPE. This requires us to compute the twist defect three point function $\braket{\Sigma_n^0\hat{D}_+ \hat{D}_+ \hat{T}_{--}}$. As described around equation \eqref{eqn:identity}, the appearence of a delta function in the $\hat{D}_+ \times \hat{D}_+$ OPE requires that the coefficient $c_n$ for $\hat{T}_{--}$ must be at least of order $(n-1)^2$ near $n =1$. We now show that this is indeed true. In the next section, we will explicitly compute the anomalous dimension of $\hat{T}_{--}$ and show that it behaves as $g_n \sim \gone(n-1) + \mathcal{O}((n-1)^2)$. We will finally show that their ratio obeys the relation
\begin{align}
\ctwo/\gone = 2\pi /S_{d-3}
\end{align}
as required by the first law of entanglement entropy.

The three point function we are after, at integer $n$, takes the form
\begin{align}\label{eqn:c_nthreepoint}
&\braket{\Sigma_n^0 \hat{T}_{--}(y') \hat{D}_+(y)\hat{D}_+(y=0)} \\ \nonumber
 &= -\oint d\bz \oint d \bw \oint \frac{du}{2\pi i u} \braket{\Sigma_n^0T_{--}(u,\bar{u}=0,y') T_{++}(z=0,\bz,y) T_{++}(w=0,\bw,0)}
\end{align}
where it is understood that all the stress tensor operators should be $\mathbb{Z}_n$ symmetrized. Our goal is now to analytically continue this expression in $n$ and then expand around $n=1$. We can turn to the previous section for this result, letting $\mathcal{O}_a = T_{++}(w= 0, \bw,0)$, $\mathcal{O}_b = T_{++}(z= 0, \bz, y)$ and $\mathcal{O}_c = T_{--}(u,\bar{u} = 0,0)$.

Just as in Section \ref{sec:proof}, a major simplification occurs for this correlator; the two displacement operators are space-like separated from each other, so they commute even upon analytic continuation. Thus, any terms with commutators between $\mathcal{O}_a$ and $\mathcal{O}_b$ in the previous section drop out.

Furthermore, the so-called ``replica diagonal" terms in the previous section will also vanish. This is because they do not contain enough $s$-integrals that produce necessary poles in $\bz$ and $\bw$. Thus, these terms vanish upon the contour integration over $\bz$ and $\bw$ in \eqref{eqn:c_nthreepoint}.

These considerations together with equation \eqref{eqn:nondiag} of the previous section make it clear that the correlator in \eqref{eqn:c_nthreepoint} vanishes up to order $(n-1)^2$. Indeed, the only surviving contribution is the second term in \eqref{eqn:nondiag}. Expanding that to second order while being careful to account for the spin of the stress tensors, we find
\begin{align}
&\braket{\Sigma_n^0 \hat{T}_{--} \hat{D}_+ \hat{D}_+}_n = \nonumber \\ 
& \frac{-(n-1)^2}{2}\oint d\bz d\bw \frac{du}{2\pi i u} \int_0^{\infty} \int_0^{\infty} d\lambda_b d\lambda_c \lambda_b^2 \lambda_c^2 \frac{\braket{ T_{++}(\bz \lambda_b,y)T_{++}(\bw \lambda_c)T_{--}(u,y') }}{(\lambda_b -1-i\epsilon)^2(\lambda_c-1+i\epsilon)^2} + \mathcal{O}((n-1)^3).
\end{align}

Rescaling $\lambda_b \to \lambda_b/\bz$ and $\lambda \to \lambda_c/\bw$, we can then expand the denominators in small $\bz,\bw$ and perform the residue projections in $\bz, \bw$ and $u.$ The final answer is the simple result
\begin{align}
&\braket{\Sigma_n^0 \hat{T}_{--} \hat{D}_+ \hat{D}_+} = 2\pi ^2(n-1)^2 \braket{\mathcal{E}_+(y)  \mathcal{E}_+(y=0)T_{--}(u=0,y')}+ \mathcal{O}((n-1)^3).
\end{align}
where $\mathcal{E}_+(y)$ is the half-averaged null energy operator
\begin{align}
\mathcal{E}_+(y) = \int_0^{\infty} d\lambda \,T_{++}(z=0,\lambda,y)
\end{align}

We now set about computing this correlator. Expanding the stress tensor three point function in a general CFT into the free field basis, we have
\begin{align}
\braket{TTT} = n_s \braket{TTT}_s + n_f \braket{TTT}_f + n_v \braket{TTT}_v
\end{align}
where $n_s, n_f$ and $n_v$ are charges characterizing the specific theory. 

One can demonstrate that the only non-vanishing contribution from these three terms is from the scalar three point function. The way to see this is as follows. The fermion term can be computed by considering a putative free Dirac fermion theory with field $\psi$. The stress tensor looks like $T_{++} \sim \bar{\psi} \Gamma_+ \partial_+ \psi$. Then we can compute the $\braket{TTT}$ three point function via Wick contractions. There will always be at least one Wick contraction between operators in each $T_{++}$. The kinematics of these operators ensure that such a contraction vanishes because they are both on the same null plane.\footnote{Actually these contractions will be proportional to a delta function $\delta^{d-2}(y)$ but we are assuming the three stress tensors sit at different $y$'s. }

The same argument can be made for the vector fields. In fact, the \emph{only} reason that the scalar contribution doesn't vanish is because of the presence of a total derivative term in the conformal stress tensor, namely $T_{++} \supset  -\frac{d-2}{4(d-1)} \partial_+^2 \col \phi^2 \col$. One can then show that the only non-vanishing term is
\begin{align}
&\braket{\mathcal{E}_+(y)\mathcal{E}_+(0)T_{--}(y')  } = \frac{4n_s (d-2)}{(d-1)^3} \frac{1}{|y|^{d-2}|y'|^{2d}}.
\end{align}
Dividing by the two point function $\braket{T_{++}(0) T_{--}(y')} = \frac{c_T}{4|y'|^{2d}}$, we find
\begin{align}\label{eqn:\ctwo}
\ctwo = \frac{32 \pi^2 n_s (d-2)}{c_T (d-1)^3}.
\end{align}
We now turn to computing the anomalous dimension $\gone$ for the stress tensor operator $\hat{T}$ on the defect.

\section{Explicit Calculation of $\gamma^{(1)}$}\label{gn}
In this section, we will follow the steps laid out in \cite{Balakrishnan:2017aa} for computing the spectrum of defect operators and associated anomalous dimension induced by the bulk stress tensor. To do this, we must compute
\begin{align}
n \sum_{j=0}^{n-1} \braket{\Sigma_n^0 T_{--}(w,0,y) T_{++}(0,\bz,0)}.
\end{align}
To leading order in $n-1$ this expression takes the form of a sum of two terms, a ``modular energy" piece and a ``relative entropy" piece
\begin{align}\label{eqn:twopoint}
(\partial_n - 1)\braket{\Sigma_n^0 \hat{T}_{--} \hat{T}_{++}}\vert_{n=1}=\left (-2\pi \braket{HT_{--}(w,0,y)T_{++}(0,\bz,0)} \right. \nonumber \\
\left. - \int_0^{-\infty} d\lambda \frac{\lambda^2}{(\lambda - 1 + i\epsilon)^2}\braket{T_{--}(w,0,y)T_{++}(0,\bz \lambda, 0)}\right) 
\end{align}

We will try to extract the anomalous dimensions and spectra of operators by examining the two point function of the defect stress tensor. In this framework, the signal of an anomalous dimension is a logarithmic divergence. As explained in \cite{Balakrishnan:2017aa}, the log needs to be cutoff by $\bz w/y^2$ or $z\bw/y^2$. In fact, there will be two such logarithms that will add to make the final answer single-valued on the Euclidean section. 

 We are thus tasked with looking for all of the terms containing $\log$ divergences in \eqref{eqn:twopoint}. Since the modular Hamiltonian is just a local integral of the stress tensor 
 
 \begin{align}
 H = \int d^{d-2}y' \int_0^{\infty} dx^+ x^+ T_{++}(x^-=0,x^+,y')
 \end{align}
then the first term on the r.h.s. of \eqref{eqn:twopoint} is a stress tensor three point function. Following the method of the previous section, we can then break up \eqref{eqn:twopoint} into the free field basis. This determines both terms on the r.h.s of $\eqref{eqn:twopoint}$ in terms of charges $n_s, n_f$ and $n_v$. This allows us to instead compute the answer in a theory of free massless scalars, fermions and vectors. While this might seem like three times the work, it actually illuminates why $g_n$ is only dependent on $n_s$. We start by examining the case of a free scalar and will see why the free fermion and free vector terms do not contribute to $g_n$.

\subsubsection*{Spectrum induced by free scalar}
This spectrum of $\phi(z,\bz, y)$ was analyzed in \cite{Bianchi_2016}. The authors found that the leading twist defect primaries are all twist one (in $d=4$) and have dimension independent of $n$. As noted in Appendix C of that work, this can be understood in any dimension from the fact that $\phi$ is annihilated by the bulk Laplacian. This constraint - for defect primaries - enforces holomorphicity in $z, \bz$ of the bulk-defect OPE which translates to a lack of anomalous dimensions. For free fermions and vectors, the same argument goes through since their two point functions are also annihilated by the Laplacian.

One might be confused because the anomalous dimension for scalar operators of dimension $\Delta$ was computed in \cite{Balakrishnan:2017aa} and found to be non-zero for operators of dimension $\Delta = \frac{d-2}{2}$. This discrepancy has to do with a subtlety related to the extra boundary term in the modular Hamiltonian for free scalars. This discrepancy is related to the choice of the stress tensor - the traceless, conformal stress tensor vs. the canonical stress tensor.

The authors of \cite{Bianchi_2016} worked with \emph{canonical} free fields, for which the stress tensor is just $T^{\text{canonical}}_{++} = \partial_+ \phi \partial_+ \phi$. Indeed if one inserts the canonical stress tensor into the modular Hamiltonian in equation (3.20) of \cite{Balakrishnan:2017aa}, then the anomalous dimension vanishes. On the other hand, if one uses the conformal stress tensor, $T^{\text{conformal}}_{++} = \col \partial_+ \phi \partial_+ \phi \col - \frac{(d-2)}{4(d-1)} \partial_+^2 \col \phi^2 \col$, then anomalous dimension for $\phi$ is given by \cite{Balakrishnan:2017aa}.

This discrepancy thus amounts to a choice of the stress tensor. Note that this is special to free scalars and does not exist for free fermions and vectors since there are no dimension $d-2$ scalar primaries in these CFTs. This proves that if one works with canonical free fields, there should be no anomalous dimension for the defect operators induced by the fundamental fields $\phi, \psi$ and $A_{\mu}$. This is enough to prove that the defect primary induced by the \emph{canonical} bulk stress tensor must also have zero anomalous dimension since this is just formed by normal-ordered products of the defect primaries induced by the bulk fundamental fields. 

\subsubsection*{Back to the stress tensor}
The upshot is that we only need to worry about the terms in \eqref{eqn:twopoint} proportional to $n_s$. Furthermore, we only need to worry about terms in the $\braket{HTT}$ term that involve the boundary term of the modular Hamiltonian. This reduces the expression down to the term
\begin{align}
\braket{HTT} \supset -\frac{(d-2)}{4(d-1)} \int d^{d-2} y \braket{\col \phi^2 \col T_{++}(0,\bz,y) T_{--}(w,0,0)}.
\end{align}
A simple calculation shows that the only contractions that give log divergences come from 
\begin{align}
\braket{HTT} &\supset \frac{n_s(d-2)^2}{4(d-1)^2} \int d^{d-2} y' \braket{\phi(0,0,y') \phi(0,0,0)} \braket{\phi(0,0,y')\partial_{\bz}^2 \phi(0,\bz,0) T_{--}(0,0,y)} \nonumber \\
& = -\frac{n_sc_{\phi \phi}^3 d(d-2)^4}{16(d-1)^3} \int d^{d-2} y' \frac{1}{|y'|^{d-2} |y-y'|^{d-2} |y|^{d+2}}.
\end{align}
This integral has two log divergences coming from $y' =0$ and $y' = y$, however they can be regulated by fixing $z,\bz$ and $w,\bw$ away from zero. The two singularities just add to make the final answer single valued under rotations by $2\pi$ about the defect as in \cite{Balakrishnan:2017aa}. We thus find 
\begin{align}
\braket{HTT} \supset & -n_s\frac{c_{\phi \phi}^3 d(d-2)^4}{32(d-1)^3} S_{d-3} \log(w\bw z \bz /|y|^4)\frac{1}{|y|^{2d}} = -\frac{2n_s (d-2)}{(d-1)^3} S_{d-3} \log(w\bw z \bz /|y|^4)\frac{1}{|y|^{2d}}.
\end{align}
Dividing by $\braket{T_{++} T_{--}}$ gives
\begin{align}
\gone = \frac{16\pi n_s (d-2)}{c_T(d-1)^3} S_{d-3}.
\end{align}
Comparing with $\eqref{eqn:\ctwo}$, we see that
\begin{align}
\frac{\ctwo}{\gone} = \frac{2\pi}{S_{d-3}}
\end{align}
as required by the first law of entanglement. 
%%%%%%%%%%%%%%%%%%%%%%%%%%%%%%%
%%%%%%%%%%%%%%%%%%%%%%%%%%%%%%+

%%%%%%%%%%%%%%%%%%%%%%%%%%%%%%%%%%%%%%%%%%
\section{Calculating $\mathcal{F}_n$} \label{app:calcFn}

At first glance, $\mathcal{F}_n$ seems difficult to calculate; we would like a method to compute this correlation function at leading order in $n-1$ without having to analytically continue a $\mathbb{Z}_n$ symmetrized four point function. The method for analytic continuation is detailed in Appendix \ref{sec:multi-rep}.

As detailed in Appendix \ref{sec:multi-rep}, part of what makes the analytic continuation in $n$ difficult is the analytic structure (branch cuts) due to various operators becoming null separated from each other in Lorentzian signature. One might naively worry that we have to track this for four operators in the four point function $\mathcal{F}_n$. 

We will leverage the fact that the two stress tensors in $\hat{D}_+(y_1)$ and $\hat{D}_+(y_2)$ are in the lightcone limit with respect to the defect since 
\begin{align}\label{eqn:D+}
\hat{D}_+(y_1) = \lim_{|z| \to 0} i\oint d\bz \sum_{j=0}^{n-1} \ T_{++}^{(j)}(z=0,\bz,y_1).
\end{align}
Thus, the stress tensors at $y_1$ and $y_2$ commute with each other even after a finite amount of boost. This means that these two operators do not see each other in the analytic continuation. In other words, the analytic structure for each of these operators is just that of a $\mathbb{Z}_n$ symmetrised \emph{three} point function. This was computed in Appendix \ref{sec:multi-rep}.

We can thus jump straight to \eqref{eqn:correction} but now with two $\mathcal{O}_b$ operators. The final replica four point function assuming $[\mathcal{O}_{b_1},\mathcal{O}_{b_2}] = 0$ is given by\footnote{We have dropped the so-called ``replica diagonal" terms in \eqref{eqn:correction} since they will drop out of the final answer after the residue projection in \eqref{eqn:D+}.}
\begin{align}\label{eqn:correctionDDDD}
&\frac{(n-1)}{8\pi^2}\int_{-\infty}^{\infty} ds_c ds_{b_1}ds_{b_2} \int_0^{s_c} dt_{b_1}dt_{b_2} \frac{\braket{[\mathcal{O}_{b_2}(-is_{b_2}-it_{b_2}),[\mathcal{O}_{b_1}(-is_{b_1}-it_{b_1}),\mathcal{O}_a(0)]]\mathcal{O}_c(-is_c+\tau_{ca})}_1}{64\sinh^2((s_{b_1} - i\tau_{b_1a})\sinh^2((s_{b_2} - i\tau_{b_2a})/2)\sinh^2((s_c - i\epsilon)/2)} \nonumber \\
& +\mathcal{O}((n-1)^2).
\end{align} 

To make contact with $\mathcal{F}_n$, we assign
\begin{align}
&\mathcal{O}_{b_1}(-is_1) = \lim_{|z| \to 0} i\oint d\bz\, e^{2s_1-2i\tau_{b_1a}}T_{++}(x^-=0,x^+=r_{\bz} e^{s_1},y_1) \nonumber \\
&\mathcal{O}_{b_2}(-is_2) = \lim_{|w| \to 0} i\oint d\bw\, e^{2s_2-2i\tau_{b_2a}}T_{++}(x^-=0,x^+=r_{\bw} e^{s_2},y_2) \nonumber \\
&\mathcal{O}_{c}(-is_c) = \lim_{|u| \to 0} i\oint du\, e^{-2s_c+2i\tau_{ca}}T_{--}(x^-=-r_u e^{-s_c},x^+=0,y_4) \nonumber \\
&\mathcal{O}_{a}(0) = \lim_{|v| \to 0} i\oint \frac{dv}{2\pi i} \, T_{--}(x^-=-r_v,x^+=0,y_3) \nonumber \\
\end{align}
with $\bz,\bw = r_{\bz,\bw}e^{i\tau_{b_1,b_2}}$ and $u,v= r_{u,v}e^{-i\tau_{a,c}}$. The funny factors of $e^{2s - 2i \tau}$ are to account for the spin of the stress tensor. 

Shifting $s_{b_{1,2}} \to s_{b_{1,2}} -t_{b_{1,2}} -\log(r_{1,2})$ and moving to null coordinates $\lambda = e^s$, we find the expression
\begin{align}\label{eqn:correctionDDDD2}
&\mathcal{F}_n = \lim_{|z|,|w|,|u|,|v| \to 0} \oint d\bz \, d\bw \, du \, dv\, \times  \nonumber \\
& \frac{(n-1)}{8\pi^2}\int_{-\infty}^{\infty} ds_c\int_{0}^{\infty} \frac{d\lambda_{b_{1,2}}\, \lambda^2_{b_1}\lambda^2_{b_2}}{\bz^3 \bw^3} \int_0^{s_c} dt_{b_1}dt_{b_2} e^{-s_c} e^{-t_{b_1} - t_{b_2}}e^{6i\tau_a} \ \times \nonumber \\
& \frac{\Braket{[T_{++}(x^+=\lambda_{b_1}),[T_{++}(x^+ = \lambda_{b_2}),T_{--}(x^-=-r_v)]]T_{--}(x^- = -r_ue^{-s_c-i\tau_{ca}})}_1}{\left(\frac{\lambda_{b_1}e^{i\tau_a}}{\bz e^{t_{b_1}}}- 1\right)^2 \left(\frac{\lambda_{b_2}e^{i\tau_a}}{\bw e^{t_{b_2}} } - 1\right)^2\left(e^{s_c-i\epsilon}-1\right)^2} .
\end{align} 

The first line in \eqref{eqn:correctionDDDD2} comes from the residue projections in the definitions of the displacement operators. Expanding the integrand at small $|\bz|$ and $|\bw|$, we can perform the residue integrals over $\bz$ and $\bw$ leaving us with 

\begin{align}\label{eqn:correctionDDDD3}
&\mathcal{F}_n = \lim_{|u|,|v| \to 0} \oint du \, dv\, \times  \nonumber \\
& \frac{1-n}{2}\int_{-\infty}^{\infty} ds_c \int_0^{s_c} dt_{b_1}dt_{b_2} e^{-s_c+2i\tau_a}e^{t_{b_1} + t_{b_2}} 
\frac{\Braket{[\mathcal{E}_+(y_1),[\mathcal{E}_+(y_2),T_{--}(x^-=-r_v)]]T_{--}(x^- = -ue^{-s_c+i\tau_a})}_1}{\left(e^{s_c-i\epsilon}-1\right)^2}
\end{align} 
where $\mathcal{E}_+(y_1)$ is a half-averaged null energy operator, $\displaystyle{\int_0^{\infty} dx^+ T_{++}(x^+)}$.

We can now do the $t_{b_1}$ and $t_{b_2}$ integrals which produce two factors of $e^{s_c}-1$ precisely cancelling the denominator. Note that a similar cancellation occurred in equation \eqref{eqn:sreldp}. We can then replace commutators of half-averaged null energy operators with commutators of full averaged null energy operators. Using the fact that $\hat{\mathcal{E}}_+\ket{\Omega} = 0$, we are left with the expression 
\begin{align}\label{eqn:DDDDfinal}
&\mathcal{F}_n = \lim_{|v|,|u| \to 0} \oint du dv\, \times  \nonumber \\
& \frac{(1-n)}{2}\int_{-\infty}^{\infty} ds_c\, e^{-s_c+2i\tau_a}\Braket{T_{--}(x^-=-r_v,x^+=0,y_3) \hat{\mathcal{E}}_+(y_1) \hat{\mathcal{E}}_+(y_2)T_{--}(x^- =-u e^{-s_c+i\tau_a},x^+=0,y_4)}_1.
\end{align} 
Using boost invariance, we can also write this as 
\begin{align}\label{eqn:DDDDFINAL}
&\mathcal{F}_n =4 \pi^2 (n-1)\int_{-\infty}^{\infty} ds_c\, e^{-s_c}\Braket{T_{--}(x^-=-1,x^+=0,y_3) \hat{\mathcal{E}}_+(y_1) \hat{\mathcal{E}}_+(y_2)T_{--}(x^- =- e^{-s_c},x^+=0,y_4)}_1
\end{align} 
where we have performed the projection over $v,u$. 

This is precisely the formula we were after. From here, one can just insert the $\hat{\mathcal{E}}_+ \times \hat{\mathcal{E}}_+$ OPE as described in the main text.

\section{Free Field Theories and Null Quantization}\label{sec:freedim}\label{sec:null}

In this section we review the basics of null quantization (see \cite{Wall:2010cj, Bousso:2015wca}). We then show that our computations in Section \ref{sec:nearvacuum} can reproduce the results of \cite{Bousso:2015wca}. In free (and super-renormalizable) quantum field theories, one can evolve the algebra of operators on some space-like slice up to the null plane $x_{-}=0$ and quantize using the null generator $P_+=\displaystyle{\int d^{d-2}y~dx^{+}~T_{++}(x^{+},y)}$ as the Hamiltonian. One can show that for free scalar fields, the algebra on the null plane factorizes across each null-generator (or ``pencil'') of the $x^-=0$ plane. For each pencil, the algebra $\mathcal{A}_{p_y}$ is just the algebra associated to a 1+1-d chiral CFT. Accordingly, the vacuum state factorizes as an infinite tensor product of $1+1$-d chiral CFT vacua:
\begin{align}
\ket{\Omega} = \bigotimes_y\ket{\Omega}^{p_y}
\end{align}
where $\ket{0}_{p_y}$ is the vacuum for the chiral $1+1$-d CFT living on the pencil at transverse coordinate $y$.

Thus, if we trace out everything to the past of some (possibly wiggly) cut of the null plane defined by $x^{+} = X^{+}(y)$, we will be left with an infinite product of reduced vacuum density matrices for a $1+1$-d CFT on the pencil
\begin{align}
\sigma_{X^{+}(y)} = \bigotimes_y \sigma^{p_y}_{x^{+} >X^{+}(y)}.
\end{align}

As discussed in \cite{Bousso:2015wca}, a general excited state on the null plane $\ket{\Psi}$ can also be expanded in the small transverse size of $\mathcal{A}$ of a given pencil. For any $p_y$, the full reduced density matrix above some cut of the null plane takes the form
\begin{align}\label{eqn:state}
 &\rho = \sigma^{p_y}_{X^{+}(y)}\otimes \rho^{(0)}_{\text{aux}} + \mathcal{A}^{1/2} \sum_{ij} \sigma^{p_y}_{X^{+}(y)} \int dr d\theta f_{ij}(r,\theta) \partial \mathcal{\phi}(r e^{i\theta}) \otimes E_{ij}(\theta)
\end{align}

where $\partial \phi$ is an operator acting on the pencil Hilbert space and $E_{ij}(\theta) = e^{\theta(K_i- K_j)}\ket{i} \bra{j}$, with $\ket{i}$ eigenvectors for the auxiliary modular Hamiltonian, $K_{\text{aux}}$. Note that $E_{ij}$ parameterizes our ignorance about the rest of the state on the null plane which is not necessarily the vacuum. 

As a consistency check of \eqref{eqn:sreldp}, we now demonstrate agreement with the result of \cite{Bousso:2015wca}. In null quantization, the delta function piece of the shape deformation corresponds to a shape deformation of the pencil while keeping the auxiliary system fixed. Note that the ansatz \ref{eqn:state} is analogous to the $\lambda$ expansion in Section \ref{sec:nearvacuum} even though we are now considering a general excited state
\begin{align}
\rho =\sigma+ \mathcal{A}^{1/2} \delta \rho + \mathcal{O}(\mathcal{A}).
\end{align}

\begin{figure}
	\centering
	\includegraphics[scale=.75]{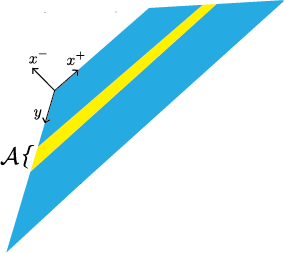}
	\caption{The Hilbert space on a null hypersurface of a free (or superrenormalizable) quantum field theory factorizes across narrow pencils of width $\mathcal{A}$. One pencil is shown above in yellow. The neighboring pencils then can be thought of as an auxiliary system (shown in blue). In the vacuum, the state between the pencil and the auxiliary system factorizes, but in an excited state there could be nontrivial entanglement between the two systems.}
	\label{fig:pencil}
\end{figure}
We now just plug in our expression of $\delta \rho$ into \eqref{relativeentropy} and find that the relative entropy second variation is
\begin{align}\label{eqn:Srel''}
\frac{d^{2}}{d{X^{+}(y)}^2} S_{\text{rel}}(\rho|\rho_{0}) 
= \frac{1}{2} \sum_{ij} \int &\int (dr d\theta)_{1}(dr d\theta)_{2} (f_{ij}(r,\theta))_{1} (f_{ji}(r,\theta))_{2} \nonumber\\
&\int ds ~e^{s}\langle (\partial \phi)_{1} \mathcal{E}_{+} \mathcal{E}_{+} (\partial \phi)_{2}(s) \rangle_{\text{p}} \langle E_{ij}(\theta_{1}) E_{ji}(\theta_{2}-i s)  \rangle_{\text{aux}}.
\end{align}

Now on the pencil, $\mathcal{E}_+$ is the translation generator so we can use the commutator $ i [ \mathcal{E}_{+}, \partial \phi] = \partial^{2} \phi$ and the fact that $\mathcal{E}_{+} \ket{0} = 0$ to get
\begin{align}
\frac{d^{2}}{d{X^{+}(y)}^2} S_{\text{rel}}(\rho|\rho_{0}) 
=\frac{1}{2}\sum_{ij} \int &\int (dr d\theta)_{1}(dr d\theta)_{2} (f_{ij}(r,\theta))_{1} (f_{ji}(r,\theta))_{2} \nonumber\\
&\int ds e^{s} \langle (\partial^3 \phi)_{1}(\partial \phi)_{2}(s) \rangle_{\text{p}} \langle E_{ij}(\theta_{1}) E_{ji}(\theta_{2}-i s)  \rangle_{\text{aux}}.
\end{align}
Using the chiral two-point function we have 
\begin{align}
 \langle (\partial^3 \phi)_{1}(\partial \phi)_{2}(s) \rangle_{\text{p}} = \frac{e^{s}}{(r_1e^{i\theta_1} -r_2e^{i\theta_2 + s})^4}.
\end{align}
Moreover, the auxiliary correlator is given by 
\begin{align}
\langle E_{ij}(\theta_1)E_{ji}(\theta_2-is) \rangle = e^{-2\pi K_i}e^{\nu_{ij}(\theta_1 - \theta_2 + is)}, \ \nu_{ij} = K_i - K_j
\end{align}

We now shift the integration contour by $s\rightarrow s +i(\theta_1-\theta_2)+i\pi + \log(r_1/r_2)$. Putting this all together we are left with evaluating 
\begin{align}
e^{-\pi (K_i + K_j)}e^{-2i(\theta_1 + \theta_2)} \left(\frac{r_1}{r_2}\right)^{i\nu_{ij}} \frac{1}{(r_1 r_2)^2}\int_{-\infty}^{\infty}ds\frac{e^{is\nu_{ij}}e^{2s}}{(1+ e^{s})^4}.
\end{align}
The $\theta$ integrals project us onto the $m=2$ Fourier modes of $f_{ij}$, $f^{(m=2)}_{ij}(r)$, and we find the final answer
\begin{align}\label{eqn:freeQ}
\frac{d^{2}}{d{X^{+}(y)}^2} S_{\text{rel}}(\rho|\rho_{0})  = \frac{1}{2}\sum_{ij}|F_{ij}^{(2)}|^2e^{-\pi (K_i + K_j)}g(\nu_{ij})
\end{align} 
where 
\begin{align}
F_{ij}^{(m)} = \int \frac{dr}{r^m}r^{i\nu_{ij}}f_{ij}^{(m)}(r), \ g(\nu) = \frac{\pi \nu (1+\nu^2)}{\text{sinh}(\pi \nu)}.
\end{align}
This is precisely the answer that was found by different methods in \cite{Bousso:2015wca}. Note that the right hand side of \eqref{eqn:freeQ} is manifestly positive as required by the QNEC.

%%%%%%%%%%%%%%%%%%%%%%%%%%%%%%%%%%%%%%%%

%%%%%%%%%%%%%%%%%%%%%%%%%%%%%%%%%%%%%%%%

\bibliographystyle{utcaps}
\bibliography{all}

\end{document}